\documentclass[sigconf,nonacm]{acmart}

\usepackage{adjustbox}
\usepackage{algorithm}
\usepackage{algpseudocode}
\usepackage{multirow}
\usepackage{cleveref}
\usepackage{subcaption}
\usepackage{array}
\usepackage{pifont}

\algrenewcommand\algorithmicrequire{\textbf{Input:}}
\algrenewcommand\algorithmicensure{\textbf{Output:}}

\AtBeginDocument{%
  }

\setcopyright{acmlicensed}
\copyrightyear{2026}
\acmYear{2026}
\acmDOI{XXXXXXX.XXXXXXX}

\acmConference[Conference acronym 'XX]{Make sure to enter the correct
  conference title from your rights confirmation emai}{June 03--05,
  2018}{Woodstock, NY}
  
\acmISBN{978-1-4503-XXXX-X/18/06}

\begin{document}

\title{Diachronic Hypergraphs for Orchestrated Multi-Agent Multimodal Memory Curation}

\author{\texorpdfstring{%
  \textbf{Yichao Feng\textsuperscript{1,2}\footnotemark[1]\authornote{Equal contributions.}},
  \textbf{Ran Zhang\textsuperscript{1}\footnotemark[1]},
  \textbf{Haoran Luo\textsuperscript{2}\footnotemark[2]\authornote{Corresponding authors.}},
  \textbf{Zhenghong Lin\textsuperscript{2}}, \\
  \textbf{Carl Yang\textsuperscript{3}},
  \textbf{Anh Tuan Luu\textsuperscript{2}\footnotemark[2]} \\
  \textsuperscript{1}LIGHTSPEED, Singapore \quad
  \textsuperscript{2}Nanyang Technological University, Singapore \\
  \textsuperscript{3}Emory University, Atlanta, USA
  \\
  \texttt{\{yichaoafeng, shrleyzhang\}@global.tencent.com,}
  \texttt{haoran.luo@ieee.org,}
  \texttt{hongzhenglin970323@gmail.com,}
  \texttt{anhtuan.luu@ntu.edu.sg}
}{Yichao Feng, Ran Zhang, Haoran Luo, Zhenghong Lin, Carl Yang, Anh Tuan Luu}}
\renewcommand{\shortauthors}{Feng et al.}

\begin{abstract}
Multi-agent systems solve tasks through collaboration, tool use, multimodal reasoning, and orchestration, but each agent operates within a knowledge boundary defined by its observations, context, and resources. Memory must preserve and transfer evidence, role specific context, decisions, procedures, and experience across interactions, not only outcomes. Vector and graph memories flatten these structures into embeddings or dyadic traces, obscuring events involving agents, tools, documents, errors, and evidence. This limits knowledge sharing, tracing, reuse, revision, and orchestration. We present MAGE, a hypergraph based multimodal database designed as a memory engine for MAS. MAGE stores agents, messages, tools, errors, procedures, documents, entities, decisions, and evidence in a heterogeneous temporal hypergraph, preserving high order collaborative events as reusable memory. It supports decision driven updates, role aware retrieval, validation, lifecycle management, and budget bounded context packing. By delivering knowledge to agents and orchestrators, MAGE expands their knowledge boundaries without modifying the models. Experiments show MAGE outperforms on various memory baselines. Our code is available.\footnote{Github Code: \url{https://github.com/Githubuseryf/MAGE}}
\end{abstract}



\maketitle

\begin{figure}[!t]
\centering
\includegraphics[width=0.9\linewidth]{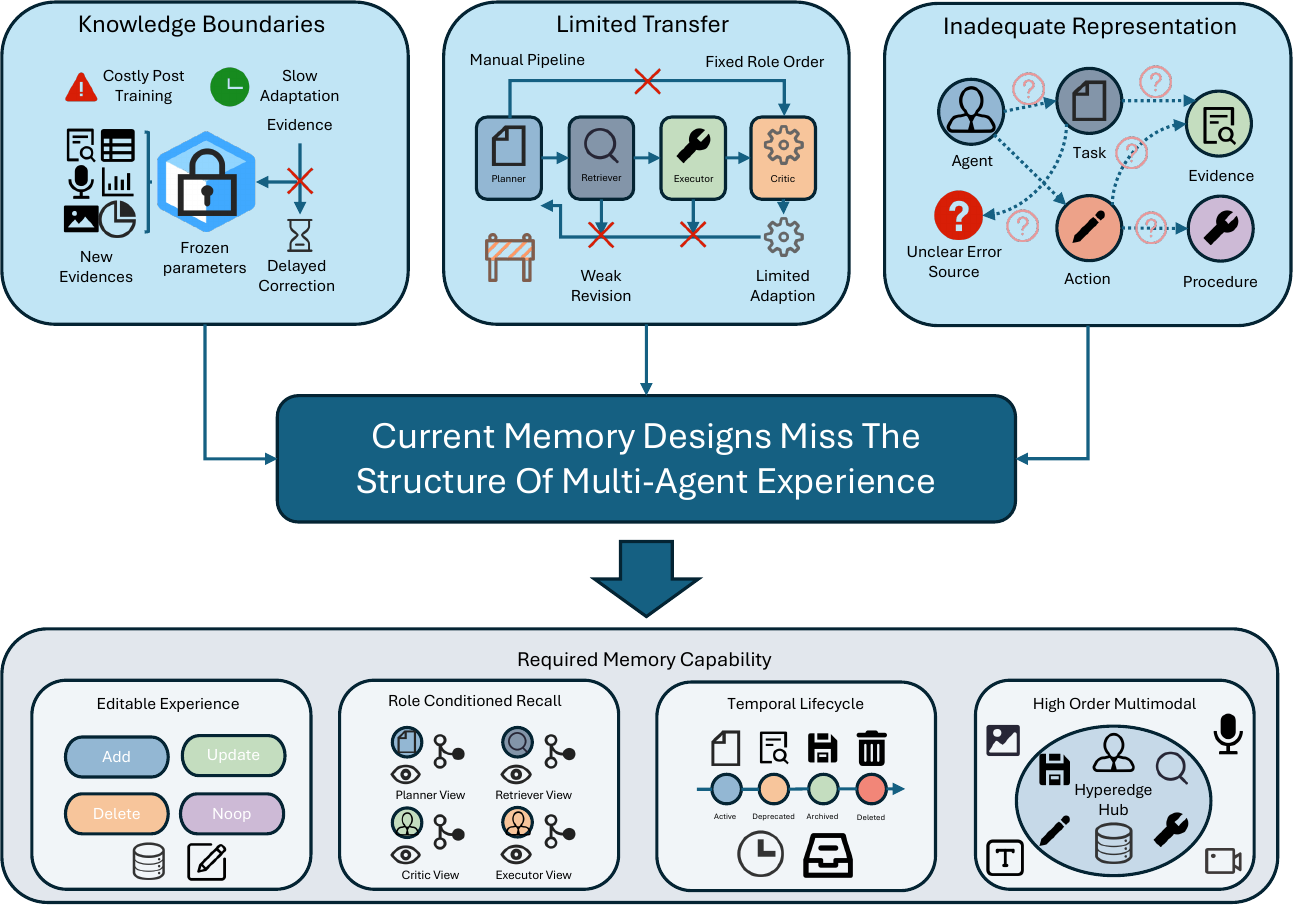}
\caption{Challenges of existing MAS memory databases.}
\label{fig:fig1}
\end{figure}

\section{Introduction}

Multi-agent systems (MAS) have emerged as a promising direction for constructing capable and adaptable AI systems \cite{dorri2018multi,ferber1999multi}. By distributing problem solving across interacting agents, MAS can enhance performance on complex tasks \cite{balaji2010introduction,chen2019control}. This design is particularly suitable for tasks requiring heterogeneous knowledge, multiple information sources, tool interactions, multimodal evidence, and iterative revision as new evidence appears. The rapid progress of large language models (LLMs)\cite{brown2020language} has further expanded the practical scope of MAS, since LLMs provide flexible agent backbones with strong language understanding, generation, coding, and reasoning abilities \cite{talebirad2023multi,li2024survey,he2025llm,han2024llm}. Yet, each agent operates within a knowledge boundary determined by its context, role, observations, tools, and histories. The collective capability of MAS therefore depends not only on individual agents and orchestration, but also on preserving and transferring knowledge across these boundaries without losing evidence, semantics, uncertainty, or provenance.

\begin{figure*}[!t]
\centering
\includegraphics[width=1\textwidth]{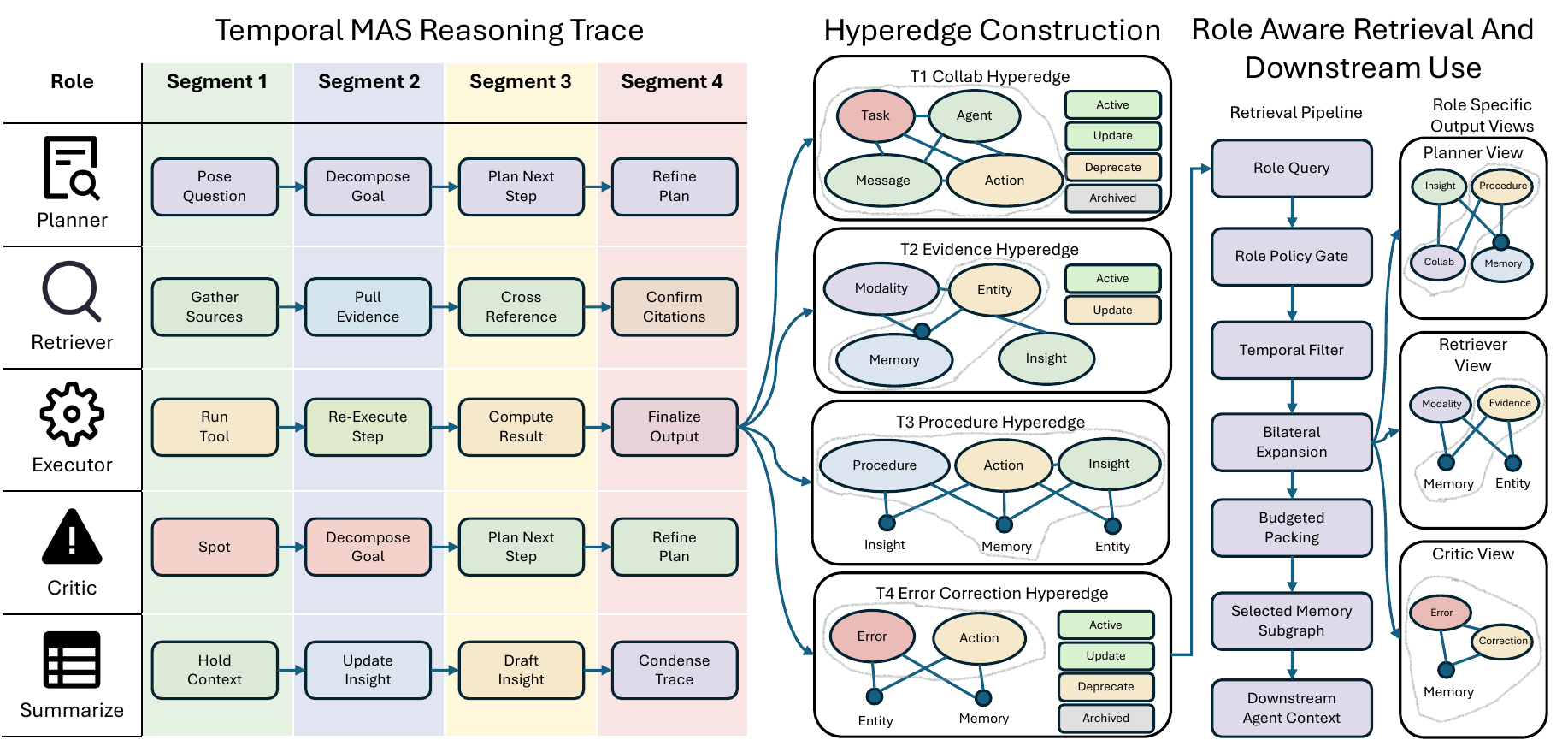}
\caption{Temporal memory construction and role aware retrieval in MAGE. MAS reasoning traces are segmented over time,
converted into various hyperedges, and later retrieved as role conditioned memory subgraphs for downstream agents.}
\label{fig:fig2}
\end{figure*}

However, MAS face three challenges, as in Fig.~\ref{fig:fig1}. First, systems lack memory for extending agent knowledge boundaries and updating knowledge over time \cite{zheng2026lifelong,wu2025memory}. Information may remain isolated within its originating agent, preventing other agents from accessing evidence, decisions, or experience when solving tasks. Incorporating it into models may require post training \cite{wu2024continual}, while repeated updates can degrade generalization and cause catastrophic forgetting \cite{luo2025empirical,li2024examining}. Second, MAS knowledge transfer is limited to textual messages, summaries, or intermediate outputs between agents and workflow stages \cite{wang-etal-2025-megaagent}. Evidence may be omitted, uncertainty compressed, role specific meanings lost, and conclusions detached from their actions or sources. These incomplete transfers can propagate errors and leave downstream agents or orchestrators unable to identify reliable, outdated, conflicting, or relevant knowledge \cite{shang2025agentsquare}. Third, memory structures remain inadequate for representing MAS knowledge \cite{yang2026graph}. Flat memories store isolated fragments, while graph or tree based memories mainly capture hierarchical or pairwise relations \cite{huang2026rethinking}. Such structures struggle to preserve high order collaborative events involving agents, roles, tools, documents, entities, intermediate states, and temporal dependencies \cite{bai2025survey}. Binary edges fragment their shared context and obscure which participants, actions, and evidence jointly produced a decision \cite{chen2026survey}.

To address these challenges, we propose \textbf{MAGE} (\textbf{M}ulti-\textbf{A}gents \textbf{G}raph \textbf{E}ngine), as shown in Fig.~\ref{fig:fig2}, a training free hypergraph memory engine that connects MAS knowledge boundaries. First, MAGE externalizes collaborative experience into a shared memory where observations, evidence, decisions, errors, and lessons can be corrected, validated, and reused without modifying LLM agents. Knowledge from one agent or stage remains accessible to agents and tasks while retaining the source, provenance, lifecycle, and evolution of transferred knowledge. Second, MAGE supports knowledge transfer through role aware retrieval. Instead of forwarding undifferentiated histories, it retrieves evidence, procedures, decisions, and experience according to the agent's role, task state, and information needs. Planners recover constraints and procedures, retrievers identify evidence requirements, executors reuse validated actions, and critics inspect sources and errors. This memory grounds orchestration when selecting agents, assigning responsibilities, and organizing interactions, making orchestration a beneficiary of transferable knowledge rather than a separate memory function. Third, MAGE introduces a heterogeneous temporal hypergraph whose multiway structure is suitable for capturing MAS interactions. An event may involve agents, roles, documents, tools, actions, states, decisions, evidence, and temporal dependencies. Pairwise edges split this event into relations, obscuring how these elements produced a result. MAGE instead preserves their many to many relations as memory units through nodes, edges, and hyperedges. Lifecycle states and provenance links retain evidence, revisions, error locations, and steps. By reconstructing collaborative episodes and retrieving role conditioned subgraphs, MAGE transfers knowledge across agent boundaries while keeping it traceable, revisable, and reusable for reasoning and orchestration.

\section{Related Work}
\label{sec:relwork}

\noindent\textbf{MAS Orchestration and Agent Memory.}
LLM based agents have been widely used for diverse tasks. ReAct\cite{yao2022react} and ToT\cite{yao2023tree} support reasoning actions and search over intermediate thoughts, while CAMEL\cite{li2023camel}, AutoGen\cite{wu2024autogen}, ChatDev\cite{qian2024chatdev}, G-memory\cite{zhang2026g}, and MetaGPT\cite{hong2024metagpt} organize MAS through role communication and SOP style workflows\cite{park2023generative}. Long term memory enables agents to reuse experience beyond a fixed context window. Agents retrieve memories by relevance, recency, and importance and synthesize reflections for planning, while Reflexion stores feedback as episodic memory\cite{shinn2023reflexion}. MemoryBank\cite{zhong2024memorybank} adds memory updates with Ebbinghaus style forgetting, and Voyager\cite{wang2023voyager} stores reusable skills for lifelong embodied learning. Recent systems such as Mem0\cite{chhikara2025mem0}, A-MEM\cite{xu2026mem}, and Zep\cite{rasmussen2025zep} explore extraction, consolidation, dynamic linking, memory evolution, temporal KG memory, and cross session knowledge maintenance across diverse application scenarios.

\begin{figure*}[!t]
\centering
\includegraphics[width=1\textwidth]{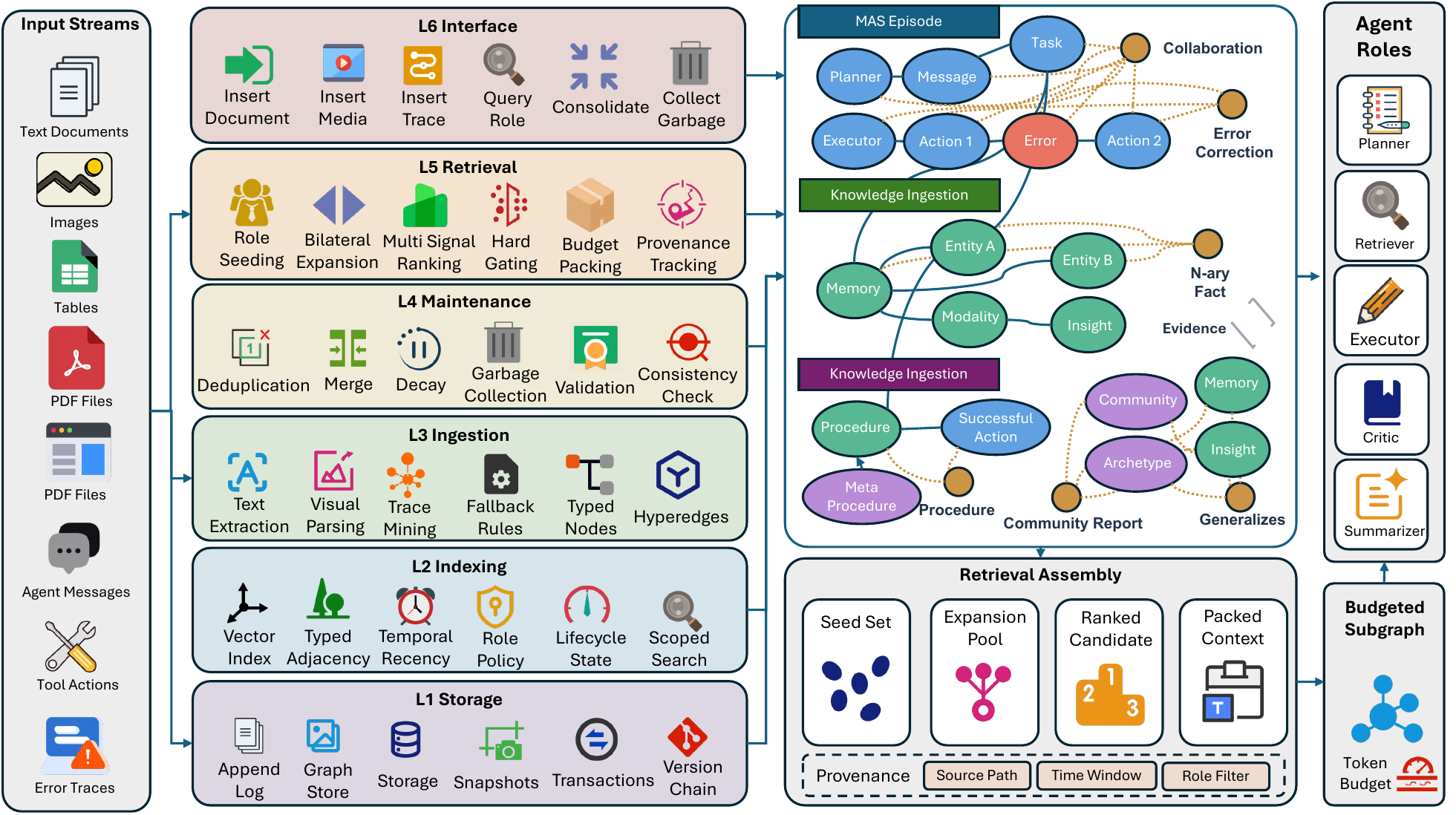}
\caption{Layered architecture of MAGE. Heterogeneous evidence streams are stored, indexed, ingested, maintained, and retrieved through six modular layers, producing role aware budgeted subgraphs from a heterogeneous temporal hypergraph.}
\label{fig:mage-arch}
\end{figure*}

\noindent\textbf{Graph and Hypergraph.}
RAG augments LLMs with external memory~\cite{lewis2020retrieval}, while graphs preserve dependencies. GraphRAG~\cite{edge2024local} builds graphs and summaries; LightRAG~\cite{guo2024lightrag} enables multilevel retrieval; RAPTOR~\cite{sarthi2024raptor} builds summary trees; and HippoRAG~\cite{gutierrez2024hipporag} uses PageRank for multi-hop retrieval. KG$^2$RAG~\cite{zhu2025knowledge} expands chunks through graph relations, whereas G-Retriever~\cite{NEURIPS2024_efaf1c97} selects subgraphs with Prize Collecting Steiner Trees. PathRAG~\cite{chen2026pathrag} retrieves relational paths, and HybGRAG~\cite{lee-etal-2025-hybgrag} combines textual and relational evidence. These remain pairwise or hierarchical. HyperGraphRAG~\cite{luo2026hypergraphrag} represents n-ary facts as hyperedges, while HyperRAG~\cite{lien2026hyperrag} retrieves n-ary relational chains. MAGE instead models mutable MAS experience, capturing how it evolves, retains provenance, and is reused across agent roles. They address distinct retrieval and memory management problems, not interchangeable systems.


\section{Preliminaries}
\label{sec:prelim}

We fix the notation that the rest of the paper relies on; design choices specific to MAGE are deferred to Sec.~\ref{sec:method}.

\paragraph{Evidence stream} An MAS $\mathcal{S}$ produces, over time, a stream
\begin{equation}
\mathcal{E}_{1:t}\;=\;\mathcal{D}_{1:t}\,\cup\,\mathcal{M}_{1:t}\,\cup\,\mathcal{T}_{1:t}
\label{eq:evidence}
\end{equation}
of textual documents, multimodal items $\sigma_j\!=\!(\mathrm{uri}_j,\mathrm{mod}_j,\tilde t_j,\mu_j)$ with $\mathrm{mod}_j\!\in\!\{$image, table, audio, video, pdf, webpage$\}$, and interaction traces
\begin{equation}
\tau_k=\bigl(q_k,\,\mathcal{A}_k,\,\pi_k=(s_{k,1},\dots,s_{k,T_k}),\,r_k\bigr),
\label{eq:trace}
\end{equation}
where $\mathcal{A}_k$ are role typed agents under $\rho:\mathcal{A}_k\!\to\!\mathcal{R}$, $\pi_k$ is a step trajectory, and $r_k$ is the final result.

\paragraph{Heterogeneous hypergraph} The data structure that absorbs this evidence is a heterogeneous graph whose nodes carry types $\mathcal{T}_V$, hyperedges types $\mathcal{T}_H$, and labelled binary edges types $\mathcal{T}_E$. A hyperedge generalizes a binary edge to a $k$ ary tuple
\begin{equation}
h=(v_{i_1},\dots,v_{i_k}),\quad v_{i_j}\!\in\!V,\;k\!\geq\!2.
\label{eq:hyperedge}
\end{equation}
For storage and indexing we use the bipartite materialization
\begin{equation}
\Phi_B(\mathcal{G})\!=\!(V\!\cup\!H,\,E\cup\{(h,v):h\!\in\!H,v\!\in\!h\}),
\label{eq:bipartite}
\end{equation}
which is bijective on $\mathcal{G}$ and lets every $k$ ary edge be served by standard bipartite or graph indexes.

\paragraph{Why hyperedges, not pairwise projections} Decomposing one $k$ ary fact into $\binom{k}{2}$ pairwise edges is lossy: it confuses co occurrence with $n$ ary co participation and erases multiplicity, an ambiguity recently formalized for hypergraph reconstruction. We therefore keep hyperedges as first class storage units.

\paragraph{Bitemporal lifecycle} Memory items must remain auditable as the world evolves, so each $x\!\in\!V\!\cup\!H$ carries a bitemporal stamp
\begin{equation}
\boldsymbol{\Theta}(x)=\bigl(\mathrm{tx}_{\mathrm{f}}(x),\mathrm{tx}_{\mathrm{t}}(x),\mathrm{val}_{\mathrm{f}}(x),\mathrm{val}_{\mathrm{t}}(x)\bigr)
\label{eq:bitemporal}
\end{equation}
separating \emph{transaction time} (when the system believed $x$) from \emph{valid time} (when $x$ holds in the world). The predicate
\begin{equation}
\mathrm{live}_t(x)\!=\!\mathbf{1}\!\bigl[\mathrm{tx}_{\mathrm{f}}(x)\!\leq\!t\!<\!\mathrm{tx}_{\mathrm{t}}(x)\bigr]\!\cdot\!\mathbf{1}\!\bigl[\mathrm{val}_{\mathrm{f}}(x)\!\leq\!t\!<\!\mathrm{val}_{\mathrm{t}}(x)\bigr]
\label{eq:live}
\end{equation}
is true exactly when both windows are open, and a lifecycle map $L\!:\!V\!\cup\!H\!\to\!\{\textsc{active},\textsc{deprecated},\textsc{archived},\textsc{deleted}\}$ records monotone descents.

\paragraph{Budget bounded query} Given a future query $q'$ under role $\rho$ and a token budget $B$, the engine returns the budget bounded subgraph
\begin{equation}
\mathcal{G}^\star_{q',\rho}\!=\!\arg\!\max_{\substack{\mathcal{G}'\subseteq\mathcal{G} \\ \mathrm{live}_t(\mathcal{G}')=1 \\ \mathrm{tok}(\mathcal{G}')\leq B}}\;\sum_{x\in\mathcal{G}'} s_\rho(x\,|\,q',\mathcal{G}),
\label{eq:objective}
\end{equation}
where $\mathrm{tok}(\cdot)$ is the rendered context token cost; this is a 0/1 budget knapsack problem with $B$ a first class constraint.

\section{Methodology}
\label{sec:method}

MAGE turns the abstract memory of Sec.~\ref{sec:prelim} into a working system.

\subsection{System Architecture}
\label{sec:method:overview}

MAGE is organized into six layers (Fig.~\ref{fig:mage-arch}), split into a \emph{data plane} and a \emph{policy plane}. 
The data plane consists of (L1) \emph{Storage}, which provides backends with snapshots and transactions; (L2) \emph{Indexing}, which maintains 
$I_t = (I^{\mathrm{vec}}_t, I^{\mathrm{adj}}_t, I^{\mathrm{tmp}}_t, I^{\mathrm{role}}_t, I^{\mathrm{lc}}_t)$ 
over semantic, structural, temporal, role, and lifecycle signals; and (L3) \emph{Ingestion}, which lifts text, multimodal items, and MAS traces into nodes and hyperedges. 
The policy plane includes (L4) \emph{Lifecycle and Consistency}, which performs deduplication, merging, decay, garbage collection, MVCC versioning, and validation; (L5) \emph{Retrieval}, which performs seeding, hypergraph expansion, ranking, and budget-bounded packing; and (L6) \emph{API}, which exposes MAS-facing insertion, retrieval, consolidation, and garbage-collection calls.

\subsection{Temporal Hypergraph Memory Model}
\label{sec:method:formal}

\textbf{Memory state.} 
At time $t$, MAGE represents a running MAS as a heterogeneous temporal hypergraph:
\begin{equation}
\mathcal{G}_t = \bigl(V_t,\,H_t,\,E_t,\,A_t,\,I_t,\,R_t,\,C_t,\,L_t\bigr).
\label{eq:hypergraph}
\end{equation}
Here, $V_t, H_t, E_t, A_t$ describe \emph{what is remembered}; $I_t, R_t$ describe \emph{how it is served}; and $C_t, L_t$ describe \emph{how it ages}.

\textbf{Graph vocabulary.} 
The vocabulary consists of ten node types and five hyperedge types:
\begin{align}
\mathcal{T}_V = \{\;& \textsc{Task},\,\textsc{Agent},\,\textsc{Message},\,\textsc{Action},\,\textsc{Error},\nonumber\\
& \textsc{Memory},\,\textsc{Insight},\,\textsc{Procedure},\nonumber\\
& \textsc{Modality},\,\textsc{Entity}\;\}, \label{eq:tv}\\
\mathcal{T}_H = \{\;& \textsc{NaryFact},\,\textsc{Collab},\,\textsc{ErrCorr},\nonumber\\
& \textsc{Evidence},\,\textsc{Procedure}\;\}. \label{eq:th}
\end{align}
The node types capture both raw collaboration evidence and durable knowledge, including tasks, agents, messages, actions, errors, memories, insights, procedures, modality pointers, and entities. 
Hyperedges preserve high-order relations: \textsc{NaryFact} groups entities in an $n$-ary statement; \textsc{Collab} binds a task with its agents and steps; \textsc{ErrCorr} links an error to corrective actions; \textsc{Evidence} ties a source to supported memories; and \textsc{Procedure} binds a workflow to its steps. 
Relations that do not require an $n$-ary container are represented by typed binary edges such as \textsc{produces}, \textsc{supersedes}, \textsc{contradicts}, \textsc{merged\_into}, and \textsc{generalizes} for efficient retrieval.

\textbf{Indexing and lifecycle.} 
Every $x \in V_t \cup H_t$ carries a schema-validated payload $a(x) \in A_t$, an embedding $\mathbf{e}_x = f_\theta(x) \in \mathbb{R}^d$, a confidence value $\kappa(x) \in [0,1]$, and the bitemporal stamp of Eq.~\eqref{eq:bitemporal}. 
The index $I_t$ partitions embeddings by type-specific scopes, so a query under role $\rho$ only touches admissible scopes. 
The role family $R_t = \{\pi_\rho\}_{\rho \in \mathcal{R}}$ stores retrieval policies, while $C_t$ and $L_t$ govern validation and aging under bounded memory budgets.

\subsection{Write Pipeline: Ingestion and Decision}
\label{sec:method:write}

When new evidence arrives, MAGE invokes the corresponding insert API, converts the input into typed graph elements, admits valid items, and commits them atomically.

\textbf{Modality Dispatch and Extraction.} 
Let $\mathcal{D}_{1:t}$ denote ingested text documents, $\mathcal{M}_{1:t}$ multimodal items, and $\mathcal{T}_{1:t}$ MAS traces. 
The graph is assembled as:
\begin{equation}
\mathcal{G}_t = \Phi_{\mathrm{txt}}(\mathcal{D}_{1:t}) \,\cup\, \Phi_{\mathrm{mm}}(\mathcal{M}_{1:t}) \,\cup\, \Phi_{\mathrm{trc}}(\mathcal{T}_{1:t}),
\label{eq:ingest-overall}
\end{equation}
where $\Phi_{\mathrm{txt}}$, $\Phi_{\mathrm{mm}}$, and $\Phi_{\mathrm{trc}}$ are LLM-backed extractors with deterministic fallbacks. 
For a text document $d$, the text extractor decomposes $d$ into a memory node, referenced entities, and $n$-ary claims:
\begin{equation}
\begin{split}
\Phi_{\mathrm{txt}}(d) =\;& \{m_d\} \cup \{e_j\}_{j=1}^{N_e(d)}\\
&\cup\; \bigl\{(m_d, e_{i_1},\!\dots,\!e_{i_k}) : (i_1,\!\dots,\!i_k)\!\in\!\mathcal{F}_d\bigr\},
\end{split}
\label{eq:naryfact}
\end{equation}
where each claim is materialized as a \textsc{NaryFact} hyperedge. 
For a multimodal item $\sigma$, MAGE creates a modality pointer and surrogate text memory:
\begin{equation}
\Phi_{\mathrm{mm}}(\sigma) = \{m_\sigma, o_\sigma\} \cup \{(o_\sigma, m_\sigma, \textsc{btm})\} \cup \Phi_{\mathrm{txt}}(\tilde t_\sigma),
\label{eq:phi-mm}
\end{equation}
where $\tilde t_\sigma$ is the extracted textual description. This surrogate design is deliberate: the memory layer stores auditable structured representations with provenance pointers to the raw item, which keeps every retrieval decision inspectable while preserving access to the original pixels whenever a downstream reasoner needs them (Appendix~\ref{app:mm-note}). 
For a trace $\tau = (q, \mathcal{A}, \pi, r)$, $\Phi_{\mathrm{trc}}$ produces task, agent, message, action, and error nodes, together with \textsc{Collab}, \textsc{ErrCorr}, and \textsc{Procedure} hyperedges. 
It further distills traces into reusable \textsc{Memory} and \textsc{Procedure} nodes with provenance for later task reuse across future contexts.

\textbf{Four-Way Decision Policy.} 
Each new memory $m^\star$ is reconciled against existing memories to prevent unbounded growth. 
MAGE retrieves top-$K$ domain-compatible neighbors $\mathcal{C}$ and applies:
\begin{align}
\delta:\,(m^\star, \mathcal{C}) \mapsto\;& (d,\,j^\star,\,m^{\diamond}),\nonumber\\
d \in\;& \{\textsc{ADD},\,\textsc{UPDATE},\,\textsc{DELETE},\,\textsc{NOOP}\},
\label{eq:delta-sig}
\end{align}
where $d$ is the selected action, $j^\star$ is the matched neighbor, and $m^\diamond$ is the merged memory under \textsc{UPDATE}. 
The decision is made by an LLM judge or a deterministic heuristic:
\begin{equation}
\delta_h(m^\star, \mathcal{C}) = \begin{cases}
(\textsc{N},\,j^\star_{\!s},\,\bot) & s^\star \geq \tau_n \wedge \Pi = 0\\
(\textsc{D},\,j^\star_{\!s},\,\bot) & s^\star \geq \tau_c \wedge \Pi = 1\\
(\textsc{U},\,j^\star_{\!s},\,m^{\diamond}) & \tau_u \leq s^\star < \tau_n\\
(\textsc{A},\,\bot,\,\bot) & \text{otherwise},
\end{cases}
\label{eq:dec-heur}
\end{equation}
where $s^\star = \max_{m_i \in \mathcal{C}} \cos(\mathbf{e}^\star, \mathbf{e}_{m_i})$, $\Pi$ detects polarity flips, and $\tau_n,\tau_c,\tau_u$ control near-duplicate, contradiction, and partial-overlap cases during incremental memory consolidation updates.

\textbf{Graph Rewrite and Persistence.} 
The decision is materialized as:
\begin{equation}
\mathcal{G}_{t+1} \!=\! \begin{cases}
\mathcal{G}_t \cup \{m^\star\} & d\!=\!\textsc{A}\\
\mathcal{G}_t\bigl[h \!\leftarrow\! (h\!\setminus\!\{m^\star\})\!\cup\!\{m_{j^\star}\}\bigr] & d\!=\!\textsc{N}\\
\mathcal{G}_t \cup \{m^\diamond,\,(m^\diamond\!\xrightarrow{\textsc{sup}}\!m_{j^\star})\} & d\!=\!\textsc{U}\\
\mathcal{G}_t \cup \{(m^\star\!\xrightarrow{\textsc{inv}}\!m_{j^\star})\} & d\!=\!\textsc{D}.
\end{cases}
\label{eq:graph-rewrite}
\end{equation}
Under \textsc{UPDATE}, $m^\diamond$ inherits provenance and the old memory is deprecated; under \textsc{DELETE}, the contradicted memory records invalidating evidence. 
Only \textsc{Memory} nodes traverse $\delta$; entities and facts are persisted unconditionally. 
Committed items are threaded into an MVCC chain 
$x^{(0)} \xrightarrow{\textsc{supersedes}} \cdots \xrightarrow{\textsc{supersedes}} x^{(v)}$, 
supporting point-in-time reads. 
A validator enforces provenance, temporal ordering, role admissibility, lifecycle consistency, merge correctness, and bounded confidence across evolving graph states.

\subsection{Read Pipeline: Role-Aware Retrieval}
\label{sec:method:read}

Any agent can call \texttt{query\_for\_role}$(q', \rho, B)$ with query $q'$, role $\rho$, and token budget $B$. 
MAGE returns a relevant, role-admissible, and budget-bounded subgraph $\mathcal{G}^\star_{q',\rho}$ through:
\begin{equation}
q' \xrightarrow{f_\theta} \mathbf{e}_{q'} \xrightarrow{\text{seed}} \mathcal{S} \xrightarrow{\text{expand}} \mathcal{X} \xrightarrow{\text{score}} \mathcal{X}^\star \xrightarrow{\text{pack}} \mathcal{G}^\star_{q',\rho}.
\label{eq:retrieve-pipe}
\end{equation}

\textbf{Seeding and Expansion.} 
The query is embedded into $\mathbf{e}_{q'}$. 
For each role-admissible scope $\sigma \in \Sigma_\rho$, MAGE returns top-$k$ items by role-boosted cosine:
\begin{equation}
\mathcal{S} = \!\!\bigcup_{\sigma \in \Sigma_\rho}\!\mathop{\arg\!\mathrm{top}\text{-}k}_{x \in V_t^{\sigma}}\!\bigl[\mathrm{sim}(\mathbf{e}_{q'}, \mathbf{e}_x) \cdot \beta_\rho(\mathrm{type}(x))\bigr] \,\cup\, \mathcal{S}_H^\rho.
\label{eq:seed}
\end{equation}
It then expands from seeds across hyperedge membership, co-participants, and typed binary neighbors:
\begin{equation}
\begin{split}
\mathcal{X}^{(\ell+1)} =\;& \mathcal{X}^{(\ell)} \cup \bigl\{y \in N_H(x) \cup N_V(x) \cup N_E(x) :\\
& x \in \mathcal{X}^{(\ell)},\; \mathrm{type}(y) \in \pi_\rho.\mathcal{T}_{\mathrm{ok}},\; \mathrm{live}_t(y)=1 \bigr\},
\end{split}
\label{eq:expand}
\end{equation}
for $\ell = 0, \dots, L{-}1$ and $\mathcal{X}^{(0)} = \mathcal{S}$. 
This supports multi-hop paths while enforcing role admissibility and temporal liveness.

\textbf{Scoring and Packing.} 
Expanded candidates are ranked by:
\begin{equation}
\begin{split}
s_\rho(c) =\;& w_{\mathrm{sem}}\,\mathrm{sim}(\mathbf{e}_{q'}, \mathbf{e}_c) + w_{\mathrm{role}}\,\beta_\rho(c) + w_{\mathrm{conf}}\,\kappa(c)\\
& + w_{\mathrm{prov}}\phi(c) + w_{\mathrm{rec}}\mathrm{rec}_t(c) + w_{\mathrm{use}}u(c)\\
& - w_{\mathrm{lc}}\ell(c) - w_{\mathrm{tok}}\tau(c).
\end{split}
\label{eq:rank}
\end{equation}
Before packing, MAGE applies:
\[
\Gamma_\rho^t(c) = \mathrm{live}_t(c) \wedge L(c) \neq \textsc{deleted} \wedge \mathrm{type}(c) \in \pi_\rho.\mathcal{T}_{\mathrm{ok}},
\]
which prevents leakage across role boundaries. 
The gated candidates are packed into a token-limited context:
\begin{equation}
\mathcal{Y}^\star = \arg\!\max_{\mathcal{Y} \subseteq \mathcal{X}^\star}\;\sum_{c \in \mathcal{Y}} s_\rho(c) \quad\text{s.t.}\quad \sum_{c \in \mathcal{Y}} \mathrm{tok}(c) \leq B.
\label{eq:knapsack}
\end{equation}
MAGE greedily admits high-value candidates and renders them as:
\[
\mathrm{ctx}(\mathcal{G}^\star_{q',\rho})
= \bigoplus_{c \in \mathcal{Y}^\star} \psi(c),
\quad
\psi(c) = [\mathrm{type}(c)]\,\xi(c).
\]
Admitted memories update their usage count and last-used timestamp.

\subsection{Maintenance: Lifecycle and Self Evolution}
\label{sec:method:maintenance}

Between writes and reads, MAGE maintains graph quality through consolidation and garbage collection. 
Two live nodes of the same type are duplicates when:
\[
\cos(\mathbf{e}_x, \mathbf{e}_y) \geq \tau_{\mathrm{c}}.
\]
The canonical winner is:
\begin{equation}
x^\dagger = \mathrm{lex\text{-}argmax}_{z \in \{x, y\}}\bigl(\kappa(z),\,|\mathrm{src}(z)|,\,-t_{\mathrm{create}}(z)\bigr),
\label{eq:canonical}
\end{equation}
favoring higher confidence, richer provenance, and older stable nodes. 
The loser is linked through \textsc{merged\_into}, and its hyperedge references are forwarded. 
MAGE also induces \textsc{Insight}, \textsc{Community}, \textsc{Archetype}, and \textsc{MetaProcedure} nodes for coarse-grained retrieval under limited context and token budgets.

\textbf{Ebbinghaus Retention and GC.}
MAGE models retention through exponential decay whose stability grows with rehearsal:
\begin{align}
R(x,t) &= \exp\!\Bigl(\!-\frac{t - t_{\mathrm{last}}(x)}{S(x)}\Bigr),\label{eq:retention}\\
S(x) &= S_0\bigl(1 + \kappa_S \ln(1 + u(x))\bigr),\nonumber
\end{align}
where $t_{\mathrm{last}}(x)$ is the last access time, $S_0$ is the base stability, $\kappa_S$ controls rehearsal scaling, and $u(x)$ is the cumulative usage count. 
Thus frequently reused memories decay more slowly, while one-off memories become GC candidates earlier.

Each live item carries a composite value:
\begin{equation}
\begin{aligned}
\nu(x) ={}& \alpha_1\kappa(x) + \alpha_2 R(x,t) + \alpha_3 u(x) + \alpha_4 g(x) \\
&+ \alpha_5 |\mathrm{src}(x)| - \alpha_6\mathrm{age}(x)
- \alpha_7\mathrm{cnt}(x) - \alpha_8\mathrm{stor}(x).
\end{aligned}
\label{eq:value}
\end{equation}
Here, $g(x)$ denotes downstream gain from successful tasks, $|\mathrm{src}(x)|$ measures provenance support, $\mathrm{cnt}(x)$ counts contradictions from newer evidence, and $\mathrm{stor}(x)$ penalizes storage cost. 
A threshold $\theta$ drives the lifecycle ladder:
\[
\textsc{active} \to \textsc{deprecated} \to \textsc{archived} \to \textsc{deleted}.
\]
After successful tasks, memories that contributed to the accepted output receive higher $u(x)$ and $g(x)$, making useful memories harder to evict while stale or contradicted memories are gradually demoted during subsequent lifecycle management rounds.


\subsection{Orchestrated Execution}
\label{sec:method:orch}

For complex tasks, MAGE uses an orchestrated solve loop backed by the shared hypergraph. 
Given task $q$ and agent pool $\mathcal{A}$, the orchestrator generates a workflow DAG, dispatches role-specific sub-tasks, retrieves memory for each role, and revises the workflow when feedback exposes unresolved gaps. 
Corrected errors become retrievable \textsc{ErrCorr} hyperedges, while successful workflows become reusable procedural memories.

\textbf{AOV Pipeline.}
The Orchestrator receives task $q$ with Planner-role memory and emits an AOV DAG over
$\mathcal{R}=\{\textsc{Pln},\textsc{Ret},\textsc{Exe},\textsc{Cri}\}.$:
\begin{equation}
\mathcal{P} = \bigl\{p_i = (\mathrm{id}_i,\,\rho_i,\,\iota_i,\,D_i)\bigr\}_{i=1}^{N_p},
\quad D_i \subseteq \{p_1,\dots,p_{i-1}\},
\label{eq:aov-dag}
\end{equation}
where $\rho_i$ is the role, $\iota_i$ is the instruction, and $D_i$ is the dependency set. 
Each sub-step retrieves memory under its role policy.

\textbf{Topological Dispatch and Re-orchestration.}
MAGE partitions $\mathcal{P}$ into topological layers and executes independent steps concurrently. 
Each step forms a compound query, retrieves role-admissible memory, executes the sub-agent, and records the output. 
The Critic evaluates all outputs; if the verdict is \textsc{Revise}, the Orchestrator produces a residual pipeline targeting the identified gaps. 
The loop stops once the result is accepted or the maximum revision depth $d_{\max}$ is reached.

\textbf{Trace Writeback.}
After termination, the full trajectory is ingested through $\Phi_{\mathrm{trc}}$ into $\mathcal{G}_t$, making accumulated experience available to future tasks while preserving provenance.

\textbf{Complexity.}
Let $n=|V_t|$, $\Delta$ be the maximum typed degree, $|\Sigma_\rho|$ the number of admissible scopes, and $|\mathcal{S}|,|\mathcal{X}|$ the seed and expanded set sizes. 
The read cost is
\begin{equation}
T_{\mathrm{read}}
  = O\!\bigl(|\Sigma_\rho|\,k\log n\bigr)
  + O\!\bigl(L\Delta|\mathcal{S}|\bigr)
  + O\!\bigl(|\mathcal{X}|\log|\mathcal{X}|\bigr).
\label{eq:read-cost}
\end{equation}
Writes cost $O(K){+}O(\Delta)$, while maintenance costs $O(n\log n)$.

\providecommand{\best}[1]{\textbf{#1}}
\providecommand{\dgain}[1]{\textcolor{teal}{\textbf{#1}}}
\providecommand{\dloss}[1]{\textcolor{red}{#1}}
\providecommand{\nm}{--}
\providecommand{\msec}[1]{\textit{#1}}

\begin{table*}[t]
\centering
\caption{Main results for RQ1 and RQ2 across GPT-4.1-mini, Claude-4.5-Sonnet, and Qwen2.5-VL-32B-Instruct backends.}
\label{tab:exp1}
\setlength{\tabcolsep}{2pt}
\renewcommand{\arraystretch}{1.05}
\scriptsize
\begin{tabular}{l cc cc cc cc cc cc cc cc cc cc cc cc}
\toprule
& \multicolumn{4}{c}{\textit{Text-only IND}} & \multicolumn{4}{c}{\textit{Text-only OOD}} & \multicolumn{8}{c}{\textit{Multimodal IND}} & \multicolumn{8}{c}{\textit{Multimodal OOD}} \\
\cmidrule(lr){2-5}\cmidrule(lr){6-9}\cmidrule(lr){10-17}\cmidrule(lr){18-25}
Method & \multicolumn{2}{c}{HotpotQA} & \multicolumn{2}{c}{NQ-Open} & \multicolumn{2}{c}{TriviaQA} & \multicolumn{2}{c}{WebQ.} & \multicolumn{2}{c}{ChartQA} & \multicolumn{2}{c}{DocVQA} & \multicolumn{2}{c}{Infogr.} & \multicolumn{2}{c}{FinMME} & \multicolumn{2}{c}{A-OKVQA} & \multicolumn{2}{c}{SciQA} & \multicolumn{2}{c}{TextVQA} & \multicolumn{2}{c}{VizWiz} \\
\cmidrule(lr){2-3} \cmidrule(lr){4-5} \cmidrule(lr){6-7} \cmidrule(lr){8-9} \cmidrule(lr){10-11} \cmidrule(lr){12-13} \cmidrule(lr){14-15} \cmidrule(lr){16-17} \cmidrule(lr){18-19} \cmidrule(lr){20-21} \cmidrule(lr){22-23} \cmidrule(lr){24-25}
& F1 & EM & F1 & EM & F1 & EM & F1 & EM & F1 & EM & F1 & EM & F1 & EM & F1 & EM & F1 & EM & F1 & EM & F1 & EM & F1 & EM \\
\midrule
\addlinespace[2pt]
\multicolumn{25}{c}{\textit{\textbf{GPT-4.1-mini results}}} \\
\addlinespace[1pt]
\midrule
\textbf{MAGE} & \best{51.08} & \best{41.67} & \best{54.70} & \best{38.54} & \best{68.17} & \best{63.54} & \best{53.48} & \best{34.38} & \best{83.71} & \best{79.17} & \best{47.64} & \best{26.04} & \best{81.94} & \best{72.92} & \best{59.22} & \best{26.04} & \best{79.29} & \best{62.50} & \best{89.55} & \best{63.54} & \best{93.65} & \best{88.54} & \best{40.79} & \best{16.67} \\
\midrule
\textsc{NoMem} & 37.30 & 27.08 & 44.41 & 22.92 & 55.91 & 54.17 & 38.66 & 16.67 & 73.32 & 67.71 & 33.86 & 15.62 & 71.29 & 58.33 & 46.18 & 8.33 & 72.48 & 53.12 & 79.87 & 57.29 & 86.15 & 81.25 & 30.23 & 7.29 \\
\textsc{Mem0} & 40.12 & 29.17 & 49.36 & 29.17 & 59.67 & 46.88 & 44.22 & \underline{29.17} & 72.48 & 59.38 & 39.69 & 16.67 & 75.01 & 64.58 & 52.08 & \underline{25.00} & 70.64 & 51.04 & 77.80 & 61.46 & 83.54 & 72.92 & 32.06 & 11.46 \\
\textsc{A-MEM} & 39.36 & 34.38 & 48.76 & 28.12 & 60.03 & 50.00 & 45.64 & 25.00 & 75.20 & 70.83 & 40.59 & 20.83 & 70.19 & 57.29 & 51.29 & 13.54 & 71.99 & 59.38 & 79.59 & \underline{62.50} & 84.54 & 77.08 & 30.14 & 4.17 \\
\textsc{MemBank} & 41.73 & 34.38 & 49.43 & \underline{37.50} & 56.70 & 50.00 & 42.06 & 25.00 & 73.79 & 69.79 & 34.86 & 10.42 & 74.24 & \underline{69.79} & 52.44 & 21.88 & 71.12 & 60.42 & 81.62 & 50.00 & 80.12 & 75.00 & 31.01 & 5.21 \\
\textsc{GenAg} & 44.83 & 31.25 & 47.69 & 27.08 & 62.71 & 55.21 & 42.11 & 20.83 & \underline{81.01} & \underline{75.00} & 34.86 & 10.42 & 75.05 & 68.75 & 47.48 & 10.42 & 72.61 & 55.21 & \underline{85.35} & 54.17 & 81.55 & 72.92 & 33.27 & 12.50 \\
\textsc{Voyager} & 43.76 & 36.46 & \underline{51.40} & 36.46 & 56.22 & 42.71 & 38.58 & 19.79 & 77.38 & 68.75 & 37.53 & 13.54 & 72.39 & 60.42 & 48.42 & 19.79 & 70.30 & 58.33 & 82.54 & 56.25 & 87.28 & 78.12 & 36.06 & 14.58 \\
\textsc{Zep} & 42.73 & \underline{38.54} & 49.35 & \underline{37.50} & \underline{66.07} & \underline{62.50} & 43.46 & 19.79 & 79.35 & 72.92 & 43.16 & 21.88 & 76.29 & 67.71 & 50.42 & 15.62 & 69.66 & 48.96 & 80.79 & 60.42 & 89.64 & 82.29 & 34.03 & 12.50 \\
\textsc{G-Mem} & 45.89 & 36.46 & 48.44 & 29.17 & 58.81 & 56.25 & 46.83 & 26.04 & 79.02 & 71.88 & 39.00 & 14.58 & 74.86 & 66.67 & 53.43 & \underline{25.00} & 71.20 & \underline{61.46} & 82.44 & \underline{62.50} & 83.85 & 76.04 & \underline{37.39} & \underline{15.62} \\
\textsc{MetaGPT} & \underline{47.50} & 35.42 & 48.94 & 30.21 & 55.47 & 53.12 & \underline{49.58} & 28.12 & 76.33 & 71.88 & 41.77 & 18.75 & \underline{78.74} & 66.67 & 52.37 & 21.88 & \underline{76.89} & 56.25 & 81.05 & 59.38 & 85.18 & 73.96 & 28.86 & 7.29 \\
\textsc{ChatDev} & 40.82 & 33.33 & 50.46 & 34.38 & 60.09 & 54.17 & 39.40 & 18.75 & 77.99 & 69.79 & 33.82 & 9.38 & 77.05 & \underline{69.79} & \underline{54.32} & 19.79 & 71.54 & 53.12 & 79.95 & 47.92 & 90.15 & 82.29 & 30.30 & 6.25 \\
\textsc{MacNet} & 43.02 & 33.33 & 47.07 & 30.21 & 58.89 & 51.04 & 41.91 & 23.96 & 75.78 & 63.54 & \underline{44.06} & \underline{22.92} & 75.31 & 65.62 & 50.30 & 23.96 & 73.44 & 54.17 & 83.82 & 58.33 & \underline{90.50} & \underline{83.33} & 34.70 & 10.42 \\
\midrule
\addlinespace[2pt]
\multicolumn{25}{c}{\textit{\textbf{Claude-4.5-Sonnet results}}} \\
\addlinespace[1pt]
\midrule
\textbf{MAGE} & \best{43.31} & \best{38.54} & \best{56.73} & \best{36.46} & \best{82.43} & \best{73.96} & \best{51.11} & \best{32.29} & \best{84.15} & \best{78.12} & \best{46.46} & \best{25.00} & \best{84.00} & \best{76.04} & \best{39.91} & \best{20.83} & \best{72.64} & \best{58.33} & \best{73.28} & \best{65.62} & \best{92.55} & \best{88.54} & \best{40.84} & \best{17.71} \\
\midrule
\textsc{NoMem} & 28.41 & 21.88 & 46.10 & 27.08 & 74.10 & 60.42 & 37.51 & 17.71 & 74.56 & 68.75 & 31.20 & 7.29 & 71.08 & 59.38 & 26.52 & 6.25 & 61.46 & 44.79 & 59.60 & 33.33 & 88.21 & 79.17 & 31.09 & 9.38 \\
\textsc{Mem0} & 32.15 & 25.00 & 52.70 & \underline{35.42} & 77.24 & 71.88 & 42.00 & 27.08 & 75.63 & 64.58 & 38.19 & 15.62 & 74.81 & 65.62 & 30.47 & 3.12 & 60.27 & 46.88 & 57.10 & 33.33 & 85.22 & 81.25 & 34.74 & 13.54 \\
\textsc{A-MEM} & 31.66 & 23.96 & 50.70 & 32.29 & 76.09 & 70.83 & 46.03 & 27.08 & 76.35 & 66.67 & 39.06 & 16.67 & 76.08 & 71.88 & 29.12 & \underline{9.38} & 63.85 & 44.79 & 58.03 & 40.62 & 85.16 & 77.08 & 33.45 & 13.54 \\
\textsc{MemBank} & 32.24 & 20.83 & 51.41 & 33.33 & 75.69 & 71.88 & 40.50 & 23.96 & 76.20 & 66.67 & 36.70 & 12.50 & 74.31 & 61.46 & \underline{34.51} & 6.25 & 62.14 & 46.88 & 61.17 & 35.42 & 82.62 & 71.88 & 33.28 & 6.25 \\
\textsc{GenAg} & 34.17 & 22.92 & 49.93 & 31.25 & 79.13 & \underline{72.92} & 39.33 & 22.92 & \underline{81.25} & 73.96 & 38.82 & 17.71 & 75.71 & 64.58 & 25.46 & 1.04 & 63.34 & 42.71 & \underline{68.48} & 42.71 & 87.73 & 85.42 & 33.30 & 8.33 \\
\textsc{Voyager} & 33.24 & 18.75 & 53.73 & \underline{35.42} & 73.97 & 64.58 & 41.22 & 26.04 & 79.61 & 73.96 & 35.51 & 11.46 & 72.94 & 63.54 & 30.06 & 1.04 & 61.28 & 43.75 & 61.84 & 38.54 & 87.71 & 84.38 & 36.98 & \underline{16.67} \\
\textsc{Zep} & 34.17 & 26.04 & 52.69 & 32.29 & \underline{80.03} & \underline{72.92} & 42.02 & 20.83 & 80.28 & 75.00 & \underline{42.96} & \underline{23.96} & 77.42 & 72.92 & 30.60 & 2.08 & 59.92 & 42.71 & 58.91 & 38.54 & 89.34 & \underline{86.46} & 35.25 & 10.42 \\
\textsc{G-Mem} & 37.91 & \underline{32.29} & 49.78 & 31.25 & 76.93 & 67.71 & \underline{47.11} & 30.21 & 80.94 & \underline{76.04} & 37.17 & 13.54 & 75.22 & 67.71 & 32.57 & 4.17 & 62.94 & 43.75 & 62.11 & 39.58 & 83.99 & 80.21 & \underline{37.44} & 15.62 \\
\textsc{MetaGPT} & 35.49 & 30.21 & 51.74 & 34.38 & 73.85 & 69.79 & 46.53 & \underline{31.25} & 76.72 & 67.71 & 38.36 & 14.58 & \underline{80.20} & \underline{75.00} & 32.26 & 4.17 & \underline{69.64} & \underline{51.04} & 62.58 & 44.79 & 85.86 & 73.96 & 31.58 & 11.46 \\
\textsc{ChatDev} & \underline{40.21} & 28.12 & \underline{54.03} & 33.33 & 77.81 & 69.79 & 38.70 & 15.62 & 79.45 & 72.92 & 39.31 & 20.83 & 77.14 & 73.96 & 33.16 & 5.21 & 62.27 & 45.83 & 57.19 & 28.12 & \underline{90.25} & 85.42 & 32.12 & 8.33 \\
\textsc{MacNet} & 32.05 & 27.08 & 49.72 & 28.12 & 76.68 & 63.54 & 40.71 & 18.75 & 72.39 & 62.50 & \underline{42.96} & 22.92 & 77.61 & 71.88 & 32.64 & 5.21 & 63.80 & 47.92 & 65.40 & \underline{48.96} & 89.63 & 80.21 & 36.66 & 14.58 \\
\midrule
\addlinespace[2pt]
\multicolumn{25}{c}{\textit{\textbf{Qwen2.5-VL-32B-Instruct results}}} \\
\addlinespace[1pt]
\midrule
\textbf{MAGE} & \best{47.35} & \best{40.62} & \best{29.08} & \best{17.71} & \best{27.52} & \best{25.00} & \best{40.61} & \best{22.92} & \best{80.56} & \best{76.04} & \best{46.62} & \best{26.04} & \best{80.74} & \best{77.08} & \best{44.66} & \best{23.96} & \best{72.52} & \best{60.42} & \best{90.02} & \best{59.38} & \best{94.41} & \best{90.62} & \best{45.99} & \best{27.08} \\
\midrule
\textsc{NoMem} & 36.56 & 23.96 & 18.10 & 3.12 & 16.45 & 11.46 & 25.88 & 8.33 & 71.34 & 65.62 & 36.03 & 12.50 & 70.12 & 58.33 & 27.71 & 7.29 & 65.79 & 50.00 & 77.59 & 52.08 & 90.31 & \underline{87.50} & 35.73 & 17.71 \\
\textsc{Mem0} & 39.23 & 25.00 & 23.60 & 13.54 & 20.61 & 7.29 & 35.30 & 14.58 & 71.92 & 60.42 & 35.90 & 11.46 & 74.14 & \underline{70.83} & 37.16 & 9.38 & 61.14 & 52.08 & 75.42 & 40.62 & 86.13 & 83.33 & 38.15 & 19.79 \\
\textsc{A-MEM} & 36.37 & 22.92 & 22.93 & 2.08 & 20.81 & 10.42 & 32.00 & 9.38 & 71.48 & 63.54 & 41.91 & 18.75 & 72.50 & 61.46 & 35.59 & 15.62 & 64.09 & 47.92 & 77.62 & 55.21 & 86.63 & 76.04 & 36.77 & 18.75 \\
\textsc{MemBank} & 37.84 & 30.21 & 24.67 & 13.54 & 19.81 & 13.54 & 29.14 & 14.58 & 71.42 & 64.58 & 33.82 & 9.38 & 70.32 & 57.29 & 36.45 & 8.33 & 64.86 & \underline{56.25} & 81.16 & 51.04 & 82.34 & 71.88 & 37.99 & 21.88 \\
\textsc{GenAg} & 41.62 & 28.12 & 20.22 & 1.04 & 21.10 & 18.75 & 28.65 & 9.38 & 72.84 & 61.46 & 32.77 & 8.33 & 74.55 & 62.50 & 31.35 & 2.08 & 63.65 & 48.96 & \underline{85.72} & \underline{58.33} & 81.06 & 75.00 & 38.59 & 19.79 \\
\textsc{Voyager} & 41.52 & 32.29 & 24.64 & 7.29 & 19.12 & 8.33 & 27.36 & 10.42 & 73.43 & 60.42 & 38.66 & 19.79 & 71.64 & 62.50 & 32.03 & 3.12 & 62.88 & 45.83 & 82.15 & 52.08 & 88.87 & 84.38 & 40.38 & 20.83 \\
\textsc{Zep} & 39.35 & 26.04 & 22.28 & 4.17 & \underline{25.62} & \underline{20.83} & 33.72 & 11.46 & \underline{75.82} & \underline{70.83} & 43.11 & 19.79 & 72.43 & 59.38 & 34.91 & 12.50 & 60.80 & 41.67 & 78.67 & 55.21 & 90.15 & 86.46 & 37.96 & 21.88 \\
\textsc{G-Mem} & 42.95 & \underline{37.50} & 21.87 & 1.04 & 18.79 & 9.38 & 35.86 & 16.67 & 75.53 & 65.62 & 41.25 & 17.71 & 73.01 & 60.42 & 38.20 & 14.58 & 64.29 & 55.21 & 79.43 & 57.29 & 84.02 & 78.12 & \underline{42.29} & \underline{22.92} \\
\textsc{MetaGPT} & 41.70 & 29.17 & 21.69 & 2.08 & 16.94 & 6.25 & \underline{37.11} & \underline{17.71} & 72.65 & 59.38 & 42.91 & 23.96 & \underline{77.04} & 68.75 & 34.64 & 7.29 & \underline{70.12} & 50.00 & 81.68 & 47.92 & 82.61 & 70.83 & 36.59 & 18.75 \\
\textsc{ChatDev} & \underline{44.05} & \underline{37.50} & \underline{26.18} & \underline{14.58} & 21.27 & 15.62 & 28.45 & 5.21 & 73.56 & 61.46 & 39.92 & 21.88 & 74.49 & \underline{70.83} & \underline{39.56} & \underline{18.75} & 64.13 & 54.17 & 77.56 & 53.12 & 87.78 & 79.17 & 35.77 & 9.38 \\
\textsc{MacNet} & 40.82 & 31.25 & 21.87 & 11.46 & 19.54 & 8.33 & 31.63 & 12.50 & 72.65 & 62.50 & \underline{44.02} & \underline{25.00} & 73.75 & 63.54 & 34.90 & 13.54 & 68.95 & 52.08 & 81.28 & 56.25 & \underline{92.11} & \underline{87.50} & 39.27 & 20.83 \\
\bottomrule
\end{tabular}
\end{table*}

\section{Experiments}
\label{sec:exp}

We evaluate MAGE using six research questions (RQs).
\textbf{RQ1}: Can MAGE improve over memory baselines across $12$ datasets?
\textbf{RQ2}: Are MAGE's gains consistent across LLM sizes and model families?
\textbf{RQ3}: Does MAGE's lifecycle machinery keep the shared memory correct, fresh, and transferable under diachronic updates and long-running operation?
\textbf{RQ4}: What are the contributions of MAGE's architectural components to overall performance?
\textbf{RQ5}: Does plugging MAGE as a drop in memory component improve performance across $8$ MAS frameworks?
\textbf{RQ6}: Can MAGE maintain efficient and consistent system level data management on the live shared graph?

\subsection{Experimental Setup}
\label{sec:exp:setup}

\noindent\textbf{Datasets.}
We evaluate on $12$ QA benchmarks spanning text and multimodal regimes:
single-hop QA (TriviaQA\cite{joshi2017triviaqa}, NQ Open\cite{kwiatkowski2019natural}, WebQuestions\cite{berant2013semantic}),
multi-hop QA (HotpotQA\cite{yang2018hotpotqa}),
financial QA (FinMME\cite{luo2025finmme}),
document/figure/chart understanding (DocVQA\cite{mathew2021docvqa}, InfographicVQA\cite{mathew2022infographicvqa}, ChartQA\cite{masry-etal-2022-chartqa}),
and visual question answering (A-OKVQA\cite{schwenk2022okvqa}, ScienceQA\cite{lu2022learn}, TextVQA\cite{singh2019towards}, VizWiz\cite{gurari2018vizwiz}).

\noindent\textbf{LLM backends.}
We use three heterogeneous LLMs: a closed lightweight model (\textsc{GPT-4.1-mini}\cite{achiam2023gpt}) used as the main driver,
a closed frontier model (\textsc{Claude-4.5-Sonnet}\cite{anthropic2025sonnet}), and an open weight LLM
(\textsc{Qwen2.5-VL-32B-Instruct}\cite{qwen2025qwen25technicalreport}). All backends share the same parameters and policy so that low level backend specific implementation details do not confound the cross LLM comparison in RQ2.

\noindent\textbf{Memory baselines.}
We compare against eleven memory or memory augmented MAS systems
spanning five families: \msec{vector memory}
(Mem0\cite{chhikara2025mem0}, A-MEM\cite{xu2026mem},
MemoryBank\cite{zhong2024memorybank}), \msec{reflection based memory}
(Generative Agents\cite{park2023generative}), \msec{skill library memory}
(Voyager\cite{wang2023voyager}), \msec{temporal graph memory}
(Zep / Graphiti\cite{rasmussen2025zep}), \msec{hierarchical MAS memory}
(G-Memory\cite{zhang2026g}), and \msec{conversation and topology centric
MAS memory} (MetaGPT\cite{hong2024metagpt}, ChatDev\cite{qian2024chatdev},
MacNet\cite{qian2025scaling}).

\noindent\textbf{Metrics.}
Utility level quality is measured by Exact Match (EM) and token level
F1, both computed against the gold answer(s) after a SQuAD style
normalisation\cite{rajpurkar2016squad}. Datasets whose source documents are indexed in the shared memory graph are reported as IND, and the remaining held out datasets as OOD. The complete protocol is given in Appendix~\ref{app:protocol}, and raw per-question predictions are released.

\begin{table*}[t]
\centering
\caption{Component ablation on GPT-4.1-mini. Variants: $-$\,Hyperedge (no $n$-ary expansion), $-$\,Role (no role-aware retrieval), $-$\,Decay (no temporal decay), $-$\,Consol.\ (no consolidation), and Retr.\,only (retrieval-only).}
\label{tab:exp3}
\setlength{\tabcolsep}{2pt}
\renewcommand{\arraystretch}{1.05}
\scriptsize
\begin{tabular}{l cc cc cc cc cc cc cc cc cc cc cc cc}
\toprule
& \multicolumn{4}{c}{\textit{Text-only IND}} & \multicolumn{4}{c}{\textit{Text-only OOD}} & \multicolumn{8}{c}{\textit{Multimodal IND}} & \multicolumn{8}{c}{\textit{Multimodal OOD}} \\
\cmidrule(lr){2-5}\cmidrule(lr){6-9}\cmidrule(lr){10-17}\cmidrule(lr){18-25}
Variant & \multicolumn{2}{c}{HotpotQA} & \multicolumn{2}{c}{NQ-Open} & \multicolumn{2}{c}{TriviaQA} & \multicolumn{2}{c}{WebQ.} & \multicolumn{2}{c}{ChartQA} & \multicolumn{2}{c}{DocVQA} & \multicolumn{2}{c}{Infogr.} & \multicolumn{2}{c}{FinMME} & \multicolumn{2}{c}{A-OKVQA} & \multicolumn{2}{c}{SciQA} & \multicolumn{2}{c}{TextVQA} & \multicolumn{2}{c}{VizWiz} \\
\cmidrule(lr){2-3} \cmidrule(lr){4-5} \cmidrule(lr){6-7} \cmidrule(lr){8-9} \cmidrule(lr){10-11} \cmidrule(lr){12-13} \cmidrule(lr){14-15} \cmidrule(lr){16-17} \cmidrule(lr){18-19} \cmidrule(lr){20-21} \cmidrule(lr){22-23} \cmidrule(lr){24-25}
& F1 & EM & F1 & EM & F1 & EM & F1 & EM & F1 & EM & F1 & EM & F1 & EM & F1 & EM & F1 & EM & F1 & EM & F1 & EM & F1 & EM \\
\midrule
\textbf{MAGE-full} & \best{51.08} & \best{41.67} & \best{54.70} & \best{38.54} & \best{68.17} & \best{63.54} & \best{53.48} & \best{34.38} & \best{83.71} & \best{79.17} & \best{47.64} & \best{26.04} & \best{81.94} & \best{72.92} & \best{59.22} & \best{26.04} & \best{79.29} & \best{62.50} & \best{89.55} & \best{63.54} & \best{93.65} & \best{88.54} & \best{40.79} & \best{16.67} \\
\midrule
$-$\,Hyperedge & 48.24 & \underline{38.54} & 52.26 & 37.50 & \underline{67.52} & \underline{62.50} & 52.29 & \underline{32.29} & 77.54 & 72.92 & 44.67 & \underline{26.04} & \underline{80.00} & \underline{71.88} & 46.14 & 19.79 & 72.80 & 51.04 & 83.13 & 56.25 & \underline{91.94} & 83.33 & 38.57 & \underline{15.62} \\
$-$\,Role & \underline{48.31} & \underline{38.54} & 53.18 & \underline{38.54} & 66.27 & 60.42 & 51.94 & 31.25 & 76.02 & 71.88 & \underline{46.16} & 25.00 & 78.93 & 68.75 & 46.68 & 20.83 & 72.64 & 48.96 & 83.79 & 55.21 & 88.97 & 80.21 & 38.09 & 14.58 \\
$-$\,Decay & 47.49 & 37.50 & 52.30 & \underline{38.54} & 66.21 & 61.46 & \underline{52.30} & \underline{32.29} & 78.10 & 73.96 & 46.10 & 25.00 & 79.97 & 69.79 & 47.11 & \underline{21.88} & 74.34 & 54.17 & 80.24 & 52.08 & 90.56 & 83.33 & 38.38 & 14.58 \\
$-$\,Consol. & 47.01 & 36.46 & \underline{53.61} & 37.50 & 66.14 & 60.42 & 51.39 & 31.25 & 75.90 & 70.83 & 44.28 & 22.92 & 79.47 & 70.83 & \underline{47.45} & \underline{21.88} & 73.60 & 52.08 & 84.33 & \underline{58.33} & 91.73 & \underline{85.42} & 37.71 & 13.54 \\
Retr.\,only & 46.97 & \underline{38.54} & 50.42 & 37.50 & 62.98 & 57.29 & 50.06 & 31.25 & \underline{79.04} & \underline{75.00} & 43.77 & \underline{26.04} & 77.95 & 69.79 & 36.89 & 12.50 & \underline{76.03} & \underline{56.25} & \underline{84.52} & 58.33 & 91.08 & 83.33 & \underline{39.48} & \underline{15.62} \\
\bottomrule
\end{tabular}
\end{table*}
\begin{figure}[t]
\centering
\includegraphics[width=0.95\columnwidth]{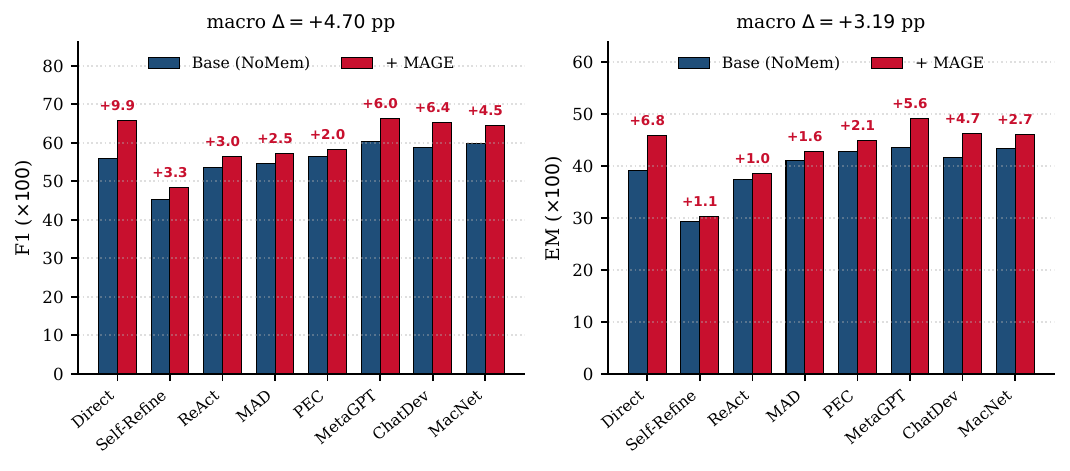}
\caption{MAGE plugged into $8$ MAS frameworks: F1 and EM of the
pure MAS baseline (blue) versus +MAGE (red).}
\label{fig:plug}
\end{figure}

\subsection{RQ1: Results with Model Matched Memory}
\label{sec:exp:e1}
RQ1 tests MAGE against baselines under the same construction and evaluation backend, isolating memory structure and retrieval. Table~\ref{tab:exp1} shows MAGE leads all 12 GPT-4.1-mini datasets, with macro F1 of 66.93 and EM of 51.13, versus 63.57 F1 for the strongest non MAGE baseline per dataset. This paired $+3.37$ margin compares each dataset's strongest baseline on exactly the same released evaluation split under identical conditions. Although several baselines are locally competitive, MAGE leads all text only and multimodal datasets. It achieves best F1 and EM on HotpotQA, NQ-Open, TriviaQA, and WebQuestions, demonstrating that the hypergraph preserves multi hop and entity centric evidence, and leads all eight multimodal datasets.


\begin{table*}[t]
\centering
\caption{MAS pluggability results on GPT-4.1-mini.}
\label{tab:rq4}
\setlength{\tabcolsep}{2pt}
\renewcommand{\arraystretch}{1.05}
\scriptsize
\resizebox{\textwidth}{!}{%
\begin{tabular}{l cc cc cc cc cc cc cc cc cc cc cc cc}
\toprule
& \multicolumn{4}{c}{\textit{Text-only IND}} & \multicolumn{4}{c}{\textit{Text-only OOD}} & \multicolumn{8}{c}{\textit{Multimodal IND}} & \multicolumn{8}{c}{\textit{Multimodal OOD}} \\
\cmidrule(lr){2-5}\cmidrule(lr){6-9}\cmidrule(lr){10-17}\cmidrule(lr){18-25}
Framework & \multicolumn{2}{c}{HotpotQA} & \multicolumn{2}{c}{NQ-Open} & \multicolumn{2}{c}{TriviaQA} & \multicolumn{2}{c}{WebQ.} & \multicolumn{2}{c}{ChartQA} & \multicolumn{2}{c}{DocVQA} & \multicolumn{2}{c}{Infogr.} & \multicolumn{2}{c}{FinMME} & \multicolumn{2}{c}{A-OKVQA} & \multicolumn{2}{c}{SciQA} & \multicolumn{2}{c}{TextVQA} & \multicolumn{2}{c}{VizWiz} \\
\cmidrule(lr){2-3} \cmidrule(lr){4-5} \cmidrule(lr){6-7} \cmidrule(lr){8-9} \cmidrule(lr){10-11} \cmidrule(lr){12-13} \cmidrule(lr){14-15} \cmidrule(lr){16-17} \cmidrule(lr){18-19} \cmidrule(lr){20-21} \cmidrule(lr){22-23} \cmidrule(lr){24-25}
& F1 & EM & F1 & EM & F1 & EM & F1 & EM & F1 & EM & F1 & EM & F1 & EM & F1 & EM & F1 & EM & F1 & EM & F1 & EM & F1 & EM \\
\midrule
Direct & 37.30 & 27.08 & 44.41 & 22.92 & 55.91 & 54.17 & 38.66 & 16.67 & 73.32 & 67.71 & 33.86 & 15.62 & 71.29 & 58.33 & 46.18 & 8.33 & 72.48 & 53.12 & 79.87 & 57.29 & 86.15 & 81.25 & 30.23 & 7.29 \\
\;\;{\footnotesize+MAGE} & \best{50.09} & \best{38.54} & \best{53.09} & \best{33.33} & \best{67.02} & \best{57.29} & \best{52.06} & \best{29.17} & \best{82.76} & \best{72.92} & \best{46.04} & \best{25.00} & \best{80.50} & \best{67.71} & \best{56.09} & \best{12.50} & \best{78.49} & \best{56.25} & \best{88.92} & \best{58.33} & \best{92.99} & \best{85.42} & \best{40.12} & \best{14.58} \\
\addlinespace[1pt]
Self-Refine & 33.27 & 22.92 & 37.40 & 21.88 & 47.04 & 36.46 & 35.80 & 19.79 & 57.32 & 45.83 & 31.26 & 14.58 & 55.43 & 41.67 & 38.95 & 14.58 & 54.46 & 36.46 & 61.06 & 36.46 & 63.85 & 52.08 & 26.18 & 8.33 \\
\;\;{\footnotesize+MAGE} & \best{37.02} & \best{23.96} & \best{39.63} & \best{22.92} & \best{49.40} & \best{37.50} & \best{38.75} & \best{20.83} & \best{60.65} & \best{46.88} & \best{34.52} & \best{15.62} & \best{59.37} & \best{42.71} & \best{42.91} & \best{15.62} & \best{57.45} & \best{37.50} & \best{64.89} & \best{37.50} & \best{67.86} & \best{53.12} & \best{29.56} & \best{9.38} \\
\addlinespace[1pt]
ReAct & 39.70 & 30.21 & 42.75 & 28.12 & 54.25 & 46.88 & 41.56 & 25.00 & 67.45 & 58.33 & 38.31 & 18.75 & 65.99 & 54.17 & 47.32 & 18.75 & 64.04 & 45.83 & 72.72 & 46.88 & 75.65 & 65.62 & 32.57 & 11.46 \\
\;\;{\footnotesize+MAGE} & \best{43.12} & \best{31.25} & \best{46.18} & \best{29.17} & \best{57.55} & \best{47.92} & \best{45.15} & \best{26.04} & \best{70.67} & \best{59.38} & \best{40.22} & \best{19.79} & \best{69.18} & \best{55.21} & \best{49.99} & \best{19.79} & \best{66.94} & \best{46.88} & \best{75.61} & \best{47.92} & \best{79.06} & \best{66.67} & \best{34.44} & \best{12.50} \\
\addlinespace[1pt]
MAD & 41.42 & 32.29 & 44.51 & 31.25 & 55.16 & 48.96 & 42.69 & 27.08 & 68.46 & 64.58 & 38.48 & 20.83 & 67.28 & 59.38 & 48.58 & 20.83 & 65.17 & 51.04 & 74.69 & 52.08 & 77.69 & 73.96 & 32.07 & 10.42 \\
\;\;{\footnotesize+MAGE} & \best{43.62} & \best{34.38} & \best{46.70} & \best{32.29} & \best{58.21} & \best{53.12} & \best{45.66} & \best{29.17} & \best{71.47} & \best{65.62} & \best{40.67} & \best{21.88} & \best{69.96} & \best{60.42} & \best{50.56} & \best{21.88} & \best{67.70} & \best{52.08} & \best{76.46} & \best{53.12} & \best{79.96} & \best{75.00} & \best{34.83} & \best{13.54} \\
\addlinespace[1pt]
PEC & 42.16 & 32.29 & 45.45 & 33.33 & 57.00 & 55.21 & 44.60 & 27.08 & 71.50 & 66.67 & 38.82 & 21.88 & 69.89 & 62.50 & 49.13 & 20.83 & 67.24 & 52.08 & 76.36 & 55.21 & 79.99 & 72.92 & 33.77 & 13.54 \\
\;\;{\footnotesize+MAGE} & \best{44.49} & \best{36.46} & \best{47.64} & \best{34.38} & \best{59.38} & \best{56.25} & \best{46.58} & \best{30.21} & \best{72.91} & \best{69.79} & \best{41.49} & \best{22.92} & \best{71.37} & \best{63.54} & \best{51.58} & \best{22.92} & \best{69.06} & \best{55.21} & \best{78.00} & \best{56.25} & \best{81.57} & \best{76.04} & \best{35.53} & \best{14.58} \\
\addlinespace[1pt]
MetaGPT & 47.50 & 35.42 & 48.94 & 30.21 & 55.47 & 53.12 & 49.58 & 28.12 & 76.33 & 71.88 & 41.77 & 18.75 & 78.74 & 66.67 & 52.37 & 21.88 & 76.89 & 56.25 & 81.05 & 59.38 & 85.18 & 73.96 & 28.86 & 7.29 \\
\;\;{\footnotesize+MAGE} & \best{50.51} & \best{40.62} & \best{54.04} & \best{36.46} & \best{67.23} & \best{60.42} & \best{53.02} & \best{32.29} & \best{82.94} & \best{76.04} & \best{47.07} & \best{25.00} & \best{81.49} & \best{72.92} & \best{58.48} & \best{23.96} & \best{78.83} & \best{60.42} & \best{89.07} & \best{61.46} & \best{92.92} & \best{84.38} & \best{39.65} & \best{15.62} \\
\addlinespace[1pt]
ChatDev & 40.82 & 33.33 & 50.46 & 34.38 & 60.09 & 54.17 & 39.40 & 18.75 & 77.99 & 69.79 & 33.82 & 9.38 & 77.05 & 69.79 & 54.32 & 19.79 & 71.54 & 53.12 & 79.95 & 47.92 & 90.15 & 82.29 & 30.30 & 6.25 \\
\;\;{\footnotesize+MAGE} & \best{49.88} & \best{35.42} & \best{53.23} & \best{36.46} & \best{66.76} & \best{58.33} & \best{50.82} & \best{29.17} & \best{81.62} & \best{70.83} & \best{46.22} & \best{23.96} & \best{80.60} & \best{70.83} & \best{58.00} & \best{22.92} & \best{76.82} & \best{54.17} & \best{87.85} & \best{57.29} & \best{92.09} & \best{83.33} & \best{38.87} & \best{12.50} \\
\addlinespace[1pt]
MacNet & 43.02 & 33.33 & 47.07 & 30.21 & 58.89 & 51.04 & 41.91 & 23.96 & 75.78 & 63.54 & 44.06 & 22.92 & 75.31 & 65.62 & 50.30 & 23.96 & 73.44 & 54.17 & 83.82 & 58.33 & 90.50 & 83.33 & 34.70 & 10.42 \\
\;\;{\footnotesize+MAGE} & \best{47.88} & \best{34.38} & \best{51.78} & \best{35.42} & \best{65.53} & \best{56.25} & \best{51.04} & \best{28.12} & \best{80.72} & \best{71.88} & \best{44.92} & \best{23.96} & \best{79.89} & \best{67.71} & \best{57.20} & \best{25.00} & \best{76.90} & \best{55.21} & \best{87.91} & \best{59.38} & \best{91.30} & \best{84.38} & \best{38.20} & \best{11.46} \\
\bottomrule
\end{tabular}%
}
\end{table*}

\begin{figure}[!t]
\centering
\includegraphics[width=0.8\linewidth]{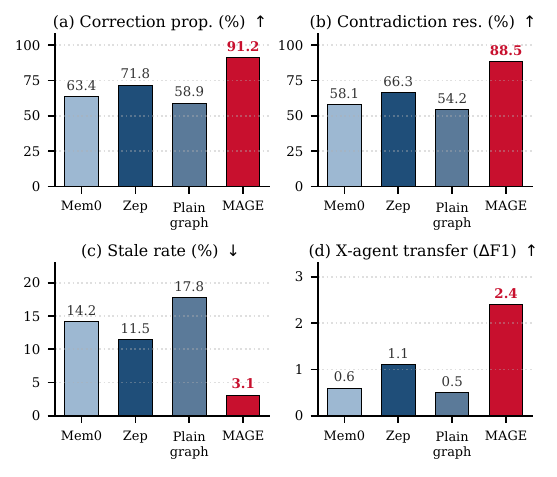}
\caption{Diachronic probes: MAGE (red)
versus baselines on \textbf{(a)} correction propagation,
\textbf{(b)} contradiction resolution, \textbf{(c)} stale rate (lower is
better), and \textbf{(d)} cross agent transfer.}
\label{fig:probes}
\end{figure}

\subsection{RQ2: Cross LLM Robustness}
\label{sec:exp:e2}
RQ2 evaluates whether MAGE remains effective across heterogeneous LLM backends with different architectures and reasoning capabilities. As shown in Table~\ref{tab:exp1}, MAGE consistently outperforms the strongest non-MAGE baseline under GPT-4.1-mini, Claude-4.5-Sonnet, and Qwen2.5-VL-32B-Instruct, with macro F1 gains of 3.37, 3.44, and 3.37 points, respectively. We note that the GPT-4.1-mini and Qwen2.5-VL-32B macro gains coincide numerically at $+3.37$; this is a coincidence of paired per question differences rather than shared predictions, since per dataset gains vary from $+1.90$ to $+5.40$ points and rank differently across backends (Table~\ref{tab:exp2}), and we release all raw predictions for per question verification. This suggests that MAGE is not tied to a specific model family or backend implementation, but reflects better organization and retrieval of external memory itself.


\subsection{RQ3: Diachronic Agent Dynamics}
\label{sec:exp:diachronic}

RQ3 asks whether shared memory remains correct, fresh, and transferable as evidence changes over time. Static QA cannot isolate these properties, so Figure~\ref{fig:probes} evaluates four controlled diachronic probes over a chronological event replay (protocol in Appendix~\ref{app:probes}). MAGE reaches 91.2\% correction propagation, 19.4 points above the strongest baseline, showing that later corrections replace superseded evidence in downstream answers. Contradiction resolution reaches 88.5\%, a 22.2 point lead, while the stale retrieval rate falls to 3.1\% from the best baseline's 11.5\%, a 73.0\% relative reduction. These two results show that explicit validity intervals, contradiction links, and lifecycle filtering prevent outdated facts from remaining equally retrievable after an update. The fourth probe tests whether memory written by one agent improves a different agent's later task. MAGE yields $+2.4$ F1 cross agent transfer, more than twice the strongest baseline's $+1.1$, indicating that role conditioned retrieval preserves reusable evidence without confining it to the originating agent. Together, the four probes support the claim that MAGE does more than improve static retrieval: it propagates revisions, resolves conflicts, suppresses stale state, and transfers experience across agent boundaries. The extended continuous replay in Appendix~\ref{app:longrunning} further shows stable performance over ten episodes rather than a one time response to injected contradictions.


\subsection{RQ4: Component Ablation}
\label{sec:exp:e3}
RQ4 examines the operational contribution of MAGE's architecture. Table~\ref{tab:exp3} shows that disabling hyperedge expansion, role aware retrieval, temporal decay, consolidation, or the combined write and maintenance path lowers performance relative to full MAGE. The per dataset losses are task dependent: higher order structure and lifecycle operations are especially important on FinMME, while role policy, decay, and consolidation make distinct contributions across multimodal tasks. A matched pairwise-projection control further isolates the schema: with nodes, embeddings, ranking, role policy, and token budget fixed, clique projection lowers macro F1 from 66.93 to 65.72 (Appendix~\ref{app:matched-control}). The retrieval only variant trails full MAGE on every dataset, showing that structured ingestion and ongoing maintenance cannot be replaced by a stronger read path alone. Appendix~\ref{app:role-sensitivity} confirms that these gains are robust to moderate changes in role weights and retrieval depth.

\begin{figure}[t]
\centering
\includegraphics[width=\linewidth]{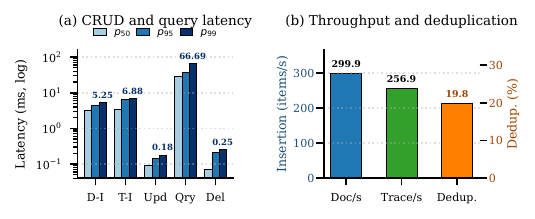}
\caption{System performance. \textbf{(a)} CRUD and query latency; D-I and T-I denote document and trace insertion. \textbf{(b)} Insertion throughput and deduplication ratio.}
\label{fig:system}
\end{figure}
\subsection{RQ5: Plug in Memory Layer Performance}
\label{sec:exp:e4}
RQ5 examines MAGE as a general external memory layer for MAS. Table~\ref{tab:rq4} and Figure~\ref{fig:plug} show gains across all eight frameworks, averaging 4.70 macro F1 and 3.19 macro EM. MAGE therefore supports single pass inference, iterative refinement, action based reasoning, debate, planning, and role structured execution without altering native coordination. \textsc{Direct} gains 9.88 F1 points, showing that structured retrieval benefits even simple pipelines. \textsc{MetaGPT} gains 6.05 points, indicating that role structured frameworks can exploit prior traces and evidence links across heterogeneous execution. These consistent improvements demonstrate that MAGE complements orchestration strategies.

\subsection{RQ6: System Level Performance}
\label{sec:exp:sys}

RQ6 evaluates MAGE as an online memory engine through efficiency and maintenance quality. Memory must support writes, updates, deletions, retrieval, and maintenance, covering access, growth, and maintenance pressure under multi agent workloads. Figure~\ref{fig:system}(a) reports MAGE's $p_{99}$ latencies: 5.25 ms for document insertion, 6.88 ms for trace insertion, 0.18 ms for update, 0.25 ms for delete, and 66.69 ms for query. MAGE maintains and retrieves hypergraph memory with limited overhead despite semantic search, role aware expansion and scoring. Figure~\ref{fig:system}(b) reports 299.9 documents/s and 256.9 traces/s, while consolidation removes 19.8\% duplicated items. These results demonstrate that MAGE sustains online operation. Graph statistics, hardware, scaling, and latency comparisons against Mem0 and Zep appear in Appendix~\ref{app:system}.

\section{Conclusion}
\label{sec:con}
We presented MAGE, an agent graph memory engine for persistent MAS. It represents experience as a heterogeneous hypergraph connecting agents, roles, evidence, entities, modalities, procedures, and traces. Role aware retrieval, temporal lifecycle management, consolidation, and budgeted packing support reasoning and data management. Across twelve text and multimodal benchmarks, MAGE outperforms memory baselines, remains robust across LLM backends, and benefits from each tested architectural path. It integrates across MAS frameworks with efficient writes, ingestion, deduplication, and retrieval. This establishes structured hypergraph memory as a basis for persistent MAS that reuse evidence, decisions, and procedural experience without modifying model parameters.

\section*{Acknowledgment}
The authors used AI-assisted tools to improve grammar and language clarity and to help debug implementation code. All AI assisted outputs were reviewed, edited, and verified by the authors, who take full responsibility for the content and originality of this work.

\bibliographystyle{ACM-Reference-Format}
\bibliography{sample-base}

@article{dorri2018multi,
  title={Multi-agent systems: A survey},
  author={Dorri, Ali and Kanhere, Salil S and Jurdak, Raja},
  journal={Ieee Access},
  volume={6},
  pages={28573--28593},
  year={2018},
  publisher={IEEE}
}

@book{ferber1999multi,
  title={Multi-agent systems: an introduction to distributed artificial intelligence},
  author={Ferber, Jacques and Weiss, Gerhard},
  volume={1},
  year={1999},
  publisher={Addison-wesley Reading}
}

@article{brown2020language,
  title={Language models are few-shot learners},
  author={Brown, Tom and Mann, Benjamin and Ryder, Nick and Subbiah, Melanie and Kaplan, Jared D and Dhariwal, Prafulla and Neelakantan, Arvind and Shyam, Pranav and Sastry, Girish and Askell, Amanda and others},
  journal={Advances in neural information processing systems},
  volume={33},
  pages={1877--1901},
  year={2020}
}

@incollection{balaji2010introduction,
  title={An introduction to multi-agent systems},
  author={Balaji, Parasumanna Gokulan and Srinivasan, Dipti},
  booktitle={Innovations in multi-agent systems and applications-1},
  pages={1--27},
  year={2010},
  publisher={Springer}
}

@article{chen2019control,
  title={On the control of multi-agent systems: A survey},
  author={Chen, Fei and Ren, Wei},
  journal={Foundations and Trends in System and Control},
  volume={6},
  number={4},
  pages={339--499},
  year={2019},
  publisher={Emerald Publishing Limited}
}

@article{talebirad2023multi,
  title={Multi-agent collaboration: Harnessing the power of intelligent llm agents},
  author={Talebirad, Yashar and Nadiri, Amirhossein},
  journal={arXiv preprint arXiv:2306.03314},
  year={2023}
}

@article{li2024survey,
  title={A survey on LLM-based multi-agent systems: workflow, infrastructure, and challenges},
  author={Li, Xinyi and Wang, Sai and Zeng, Siqi and Wu, Yu and Yang, Yi},
  journal={Vicinagearth},
  volume={1}, 
  number={1},
  pages={9},
  year={2024},
  publisher={Springer}
}

@article{he2025llm,
  title={Llm-based multi-agent systems for software engineering: Literature review, vision, and the road ahead},
  author={He, Junda and Treude, Christoph and Lo, David},
  journal={ACM Transactions on Software Engineering and Methodology},
  volume={34},
  number={5},
  pages={1--30},
  year={2025},
  publisher={ACM New York, NY}
}

@article{han2024llm,
  title={LLM multi-agent systems: Challenges and open problems},
  author={Han, Shanshan and Zhang, Qifan and Jin, Weizhao and Xu, Zhaozhuo},
  journal={arXiv preprint arXiv:2402.03578},
  year={2024}
}

@article{zheng2026lifelong,
  title={Lifelong learning of large language model based agents: A roadmap},
  author={Zheng, Junhao and Shi, Chengming and Cai, Xidi and Li, Qiuke and Zhang, Duzhen and Li, Chenxing and Yu, Dong and Ma, Qianli},
  journal={IEEE Transactions on Pattern Analysis and Machine Intelligence},
  year={2026},
  publisher={IEEE}
}

@article{wu2025memory,
  title={Memory in llm-based multi-agent systems: Mechanisms, challenges, and collective intelligence},
  author={Wu, Shanglin and Shu, Kai},
  journal={Authorea Preprints},
  year={2025},
  publisher={Authorea}
}

@article{wu2024continual,
  title={Continual learning for large language models: A survey},
  author={Wu, Tongtong and Luo, Linhao and Li, Yuan-Fang and Pan, Shirui and Vu, Thuy-Trang and Haffari, Gholamreza},
  journal={arXiv preprint arXiv:2402.01364},
  year={2024}
}

@article{luo2025empirical,
  title={An empirical study of catastrophic forgetting in large language models during continual fine-tuning},
  author={Luo, Yun and Yang, Zhen and Meng, Fandong and Li, Yafu and Zhou, Jie and Zhang, Yue},
  journal={IEEE Transactions on Audio, Speech and Language Processing},
  year={2025},
  publisher={IEEE}
}

@article{li2024examining,
  title={Examining forgetting in continual pre-training of aligned large language models},
  author={Li, Chen-An and Lee, Hung-Yi},
  journal={arXiv preprint arXiv:2401.03129},
  year={2024}
}

@inproceedings{wang-etal-2025-megaagent,
    title = "{M}ega{A}gent: A Large-Scale Autonomous {LLM}-based Multi-Agent System Without Predefined {SOP}s",
    author = "Wang, Qian  and
      Wang, Tianyu  and
      Tang, Zhenheng  and
      Li, Qinbin  and
      Chen, Nuo  and
      Liang, Jingsheng  and
      He, Bingsheng",
    editor = "Che, Wanxiang  and
      Nabende, Joyce  and
      Shutova, Ekaterina  and
      Pilehvar, Mohammad Taher",
    booktitle = "Findings of the Association for Computational Linguistics: ACL 2025",
    month = jul,
    year = "2025",
    address = "Vienna, Austria",
    publisher = "Association for Computational Linguistics",
    url = "https://aclanthology.org/2025.findings-acl.259/",
    doi = "10.18653/v1/2025.findings-acl.259",
    pages = "4998--5036",
    ISBN = "979-8-89176-256-5"
}

@inproceedings{shang2025agentsquare,
  title={Agentsquare: Automatic llm agent search in modular design space},
  author={Shang, Yu and Li, Yu and Zhao, Keyu and Ma, Likai and Liu, Jiahe and Xu, Fengli and Li, Yong},
  booktitle={International Conference on Learning Representations},
  volume={2025},
  pages={3841--3865},
  year={2025}
}

@article{yang2026graph,
  title={Graph-based Agent Memory: Taxonomy, Techniques, and Applications},
  author={Yang, Chang and Zhou, Chuang and Xiao, Yilin and Dong, Su and Zhuang, Luyao and Zhang, Yujing and Wang, Zhu and Hong, Zijin and Yuan, Zheng and Xiang, Zhishang and others},
  journal={arXiv preprint arXiv:2602.05665},
  year={2026}
}

@article{huang2026rethinking,
  title={Rethinking Memory Mechanisms of Foundation Agents in the Second Half: A Survey},
  author={Huang, Wei-Chieh and Zhang, Weizhi and Liang, Yueqing and Bei, Yuanchen and Chen, Yankai and Feng, Tao and Pan, Xinyu and Tan, Zhen and Wang, Yu and Wei, Tianxin and others},
  journal={arXiv preprint arXiv:2602.06052},
  year={2026}
}

@article{chen2026survey,
  title={A Survey of Agentic GraphRAG: From Retrieval-Augmented Generation to Graph-Native Agents},
  author={Chen, Zihan and Zheng, Lei and Zhu, Di},
  journal={Available at SSRN 6713979},
  year={2026}
}

@article{bai2025survey,
  title={Survey on AI Memory: Theories, Taxonomies, Evaluations, and Emerging Trends},
  author={BAI, TING and FAN, JIAYANG and WEN, XIAOSHUAI and KANG, JIAZHENG and LAN, HENGZHI and ZHAO, RUOCHEN and WU, PINGZHENG and ZHANG, ZEPENG and ZHONG, YUTIAN and LI, GEZI and others},
  year={2025}
}

@article{li2023camel,
  title={Camel: Communicative agents for" mind" exploration of large language model society},
  author={Li, Guohao and Hammoud, Hasan and Itani, Hani and Khizbullin, Dmitrii and Ghanem, Bernard},
  journal={Advances in neural information processing systems},
  volume={36},
  pages={51991--52008},
  year={2023}
}

@article{yao2022react,
  title={React: Synergizing reasoning and acting in language models},
  author={Yao, Shunyu and Zhao, Jeffrey and Yu, Dian and Du, Nan and Shafran, Izhak and Narasimhan, Karthik and Cao, Yuan},
  journal={arXiv preprint arXiv:2210.03629},
  year={2022}
}

@article{yao2023tree,
  title={Tree of thoughts: Deliberate problem solving with large language models},
  author={Yao, Shunyu and Yu, Dian and Zhao, Jeffrey and Shafran, Izhak and Griffiths, Tom and Cao, Yuan and Narasimhan, Karthik},
  journal={Advances in neural information processing systems},
  volume={36},
  pages={11809--11822},
  year={2023}
}

@inproceedings{wu2024autogen,
  title={Autogen: Enabling next-gen LLM applications via multi-agent conversations},
  author={Wu, Qingyun and Bansal, Gagan and Zhang, Jieyu and Wu, Yiran and Li, Beibin and Zhu, Erkang and Jiang, Li and Zhang, Xiaoyun and Zhang, Shaokun and Liu, Jiale and others},
  booktitle={First conference on language modeling},
  year={2024}
}

@inproceedings{qian2024chatdev,
  title={Chatdev: Communicative agents for software development},
  author={Qian, Chen and Liu, Wei and Liu, Hongzhang and Chen, Nuo and Dang, Yufan and Li, Jiahao and Yang, Cheng and Chen, Weize and Su, Yusheng and Cong, Xin and others},
  booktitle={Proceedings of the 62nd annual meeting of the association for computational linguistics (volume 1: Long papers)},
  pages={15174--15186},
  year={2024}
}

@inproceedings{hong2024metagpt,
  title={MetaGPT: Meta programming for a multi-agent collaborative framework},
  author={Hong, Sirui and Zhuge, Mingchen and Chen, Jonathan and Zheng, Xiawu and Cheng, Yuheng and Wang, Jinlin and Zhang, Ceyao and Yau, Steven and Lin, Zijuan and Zhou, Liyang and others},
  booktitle={International Conference on Learning Representations},
  volume={2024},
  pages={23247--23275},
  year={2024}
}

@article{zhang2026g,
  title={G-memory: Tracing hierarchical memory for multi-agent systems},
  author={Zhang, Guibin and Fu, Muxin and Wang, Kun and Wan, Frank and Yu, Miao and Yan, Shuicheng},
  journal={Advances in Neural Information Processing Systems},
  volume={38},
  pages={12988--13018},
  year={2026}
}

@inproceedings{park2023generative,
  title={Generative agents: Interactive simulacra of human behavior},
  author={Park, Joon Sung and O'Brien, Joseph and Cai, Carrie Jun and Morris, Meredith Ringel and Liang, Percy and Bernstein, Michael S},
  booktitle={Proceedings of the 36th annual acm symposium on user interface software and technology},
  pages={1--22},
  year={2023}
}

@article{shinn2023reflexion,
  title={Reflexion: Language agents with verbal reinforcement learning},
  author={Shinn, Noah and Cassano, Federico and Gopinath, Ashwin and Narasimhan, Karthik and Yao, Shunyu},
  journal={Advances in neural information processing systems},
  volume={36},
  pages={8634--8652},
  year={2023}
}

@inproceedings{zhong2024memorybank,
  title={Memorybank: Enhancing large language models with long-term memory},
  author={Zhong, Wanjun and Guo, Lianghong and Gao, Qiqi and Ye, He and Wang, Yanlin},
  booktitle={Proceedings of the AAAI conference on artificial intelligence},
  volume={38},
  number={17},
  pages={19724--19731},
  year={2024}
}

@article{wang2023voyager,
  title={Voyager: An open-ended embodied agent with large language models},
  author={Wang, Guanzhi and Xie, Yuqi and Jiang, Yunfan and Mandlekar, Ajay and Xiao, Chaowei and Zhu, Yuke and Fan, Linxi and Anandkumar, Anima},
  journal={arXiv preprint arXiv:2305.16291},
  year={2023}
}

@article{chhikara2025mem0,
  title={Mem0: Building production-ready ai agents with scalable long-term memory},
  author={Chhikara, Prateek and Khant, Dev and Aryan, Saket and Singh, Taranjeet and Yadav, Deshraj},
  journal={arXiv preprint arXiv:2504.19413},
  year={2025}
}

@article{xu2026mem,
  title={A-mem: Agentic memory for llm agents},
  author={Xu, Wujiang and Liang, Zujie and Mei, Kai and Gao, Hang and Tan, Juntao and Zhang, Yongfeng},
  journal={Advances in Neural Information Processing Systems},
  volume={38},
  pages={17577--17604},
  year={2026}
}

@article{rasmussen2025zep,
  title={Zep: a temporal knowledge graph architecture for agent memory},
  author={Rasmussen, Preston and Paliychuk, Pavlo and Beauvais, Travis and Ryan, Jack and Chalef, Daniel},
  journal={arXiv preprint arXiv:2501.13956},
  year={2025}
}

@article{lewis2020retrieval,
  title={Retrieval-augmented generation for knowledge-intensive nlp tasks},
  author={Lewis, Patrick and Perez, Ethan and Piktus, Aleksandra and Petroni, Fabio and Karpukhin, Vladimir and Goyal, Naman and K{\"u}ttler, Heinrich and Lewis, Mike and Yih, Wen-tau and Rockt{\"a}schel, Tim and others},
  journal={Advances in neural information processing systems},
  volume={33},
  pages={9459--9474},
  year={2020}
}

@article{edge2024local,
  title={From local to global: A graph rag approach to query-focused summarization},
  author={Edge, Darren and Trinh, Ha and Cheng, Newman and Bradley, Joshua and Chao, Alex and Mody, Apurva and Truitt, Steven and Metropolitansky, Dasha and Ness, Robert Osazuwa and Larson, Jonathan},
  journal={arXiv preprint arXiv:2404.16130},
  year={2024}
}

@article{guo2024lightrag,
  title={Lightrag: Simple and fast retrieval-augmented generation},
  author={Guo, Zirui and Xia, Lianghao and Yu, Yanhua and Ao, Tian and Huang, Chao},
  journal={arXiv preprint arXiv:2410.05779},
  volume={2},
  number={3},
  year={2024}
}

@inproceedings{sarthi2024raptor,
  title={Raptor: Recursive abstractive processing for tree-organized retrieval},
  author={Sarthi, Parth and Abdullah, Salman and Tuli, Aditi and Khanna, Shubh and Goldie, Anna and Manning, Christopher},
  booktitle={International Conference on Learning Representations},
  volume={2024},
  pages={32628--32649},
  year={2024}
}

@article{gutierrez2024hipporag,
  title={Hipporag: Neurobiologically inspired long-term memory for large language models},
  author={Guti{\'e}rrez, Bernal J and Shu, Yiheng and Gu, Yu and Yasunaga, Michihiro and Su, Yu},
  journal={Advances in neural information processing systems},
  volume={37},
  pages={59532--59569},
  year={2024}
}

@article{luo2026hypergraphrag,
  title={Hypergraphrag: Retrieval-augmented generation via hypergraph-structured knowledge representation},
  author={Luo, Haoran and Chen, Guanting and Zheng, Yandan and Wu, Xiaobao and Guo, Yikai and Lin, Qika and Feng, Yu and Kuang, Zemin and Song, Meina and Zhu, Yifan and others},
  journal={Advances in Neural Information Processing Systems},
  volume={38},
  pages={152206--152234},
  year={2026}
}

@inproceedings{joshi2017triviaqa,
  title={Triviaqa: A large scale distantly supervised challenge dataset for reading comprehension},
  author={Joshi, Mandar and Choi, Eunsol and Weld, Daniel S and Zettlemoyer, Luke},
  booktitle={Proceedings of the 55th Annual Meeting of the Association for Computational Linguistics (Volume 1: Long Papers)},
  pages={1601--1611},
  year={2017}
}

@article{kwiatkowski2019natural,
  title={Natural questions: a benchmark for question answering research},
  author={Kwiatkowski, Tom and Palomaki, Jennimaria and Redfield, Olivia and Collins, Michael and Parikh, Ankur and Alberti, Chris and Epstein, Danielle and Polosukhin, Illia and Devlin, Jacob and Lee, Kenton and others},
  journal={Transactions of the Association for Computational Linguistics},
  volume={7},
  pages={453--466},
  year={2019},
  publisher={MIT Press One Rogers Street, Cambridge, MA 02142-1209, USA journals-info~…}
}

@inproceedings{berant2013semantic,
  title={Semantic parsing on freebase from question-answer pairs},
  author={Berant, Jonathan and Chou, Andrew and Frostig, Roy and Liang, Percy},
  booktitle={Proceedings of the 2013 conference on empirical methods in natural language processing},
  pages={1533--1544},
  year={2013}
}

@inproceedings{yang2018hotpotqa,
  title={HotpotQA: A dataset for diverse, explainable multi-hop question answering},
  author={Yang, Zhilin and Qi, Peng and Zhang, Saizheng and Bengio, Yoshua and Cohen, William and Salakhutdinov, Ruslan and Manning, Christopher D},
  booktitle={Proceedings of the 2018 conference on empirical methods in natural language processing},
  pages={2369--2380},
  year={2018}
}

@inproceedings{luo2025finmme,
  title={Finmme: Benchmark dataset for financial multi-modal reasoning evaluation},
  author={Luo, Junyu and Kou, Zhizhuo and Yang, Liming and Luo, Xiao and Huang, Jinsheng and Xiao, Zhiping and Peng, Jingshu and Liu, Chengzhong and Ji, Jiaming and Liu, Xuanzhe and others},
  booktitle={Proceedings of the 63rd Annual Meeting of the Association for Computational Linguistics (Volume 1: Long Papers)},
  pages={29465--29489},
  year={2025}
}

@inproceedings{mathew2021docvqa,
  title={Docvqa: A dataset for vqa on document images},
  author={Mathew, Minesh and Karatzas, Dimosthenis and Jawahar, CV},
  booktitle={Proceedings of the IEEE/CVF winter conference on applications of computer vision},
  pages={2200--2209},
  year={2021}
}

@inproceedings{mathew2022infographicvqa,
  title={Infographicvqa},
  author={Mathew, Minesh and Bagal, Viraj and Tito, Rub{\`e}n and Karatzas, Dimosthenis and Valveny, Ernest and Jawahar, CV},
  booktitle={Proceedings of the IEEE/CVF Winter Conference on Applications of Computer Vision},
  pages={1697--1706},
  year={2022}
}

@inproceedings{masry-etal-2022-chartqa,
    title = "{C}hart{QA}: A Benchmark for Question Answering about Charts with Visual and Logical Reasoning",
    author = "Masry, Ahmed  and
      Long, Do Xuan  and
      Tan, Jia Qing  and
      Joty, Shafiq  and
      Hoque, Enamul",
    editor = "Muresan, Smaranda  and
      Nakov, Preslav  and
      Villavicencio, Aline",
    booktitle = "Findings of the Association for Computational Linguistics: ACL 2022",
    month = may,
    year = "2022",
    address = "Dublin, Ireland",
    publisher = "Association for Computational Linguistics",
    url = "https://aclanthology.org/2022.findings-acl.177/",
    doi = "10.18653/v1/2022.findings-acl.177",
    pages = "2263--2279"
}

@inproceedings{schwenk2022okvqa,
  title={A-okvqa: A benchmark for visual question answering using world knowledge},
  author={Schwenk, Dustin and Khandelwal, Apoorv and Clark, Christopher and Marino, Kenneth and Mottaghi, Roozbeh},
  booktitle={European conference on computer vision},
  pages={146--162},
  year={2022},
  organization={Springer}
}

@article{lu2022learn,
  title={Learn to explain: Multimodal reasoning via thought chains for science question answering},
  author={Lu, Pan and Mishra, Swaroop and Xia, Tanglin and Qiu, Liang and Chang, Kai-Wei and Zhu, Song-Chun and Tafjord, Oyvind and Clark, Peter and Kalyan, Ashwin},
  journal={Advances in neural information processing systems},
  volume={35},
  pages={2507--2521},
  year={2022}
}

@inproceedings{singh2019towards,
  title={Towards vqa models that can read},
  author={Singh, Amanpreet and Natarajan, Vivek and Shah, Meet and Jiang, Yu and Chen, Xinlei and Batra, Dhruv and Parikh, Devi and Rohrbach, Marcus},
  booktitle={Proceedings of the IEEE/CVF conference on computer vision and pattern recognition},
  pages={8317--8326},
  year={2019}
}

@inproceedings{gurari2018vizwiz,
  title={Vizwiz grand challenge: Answering visual questions from blind people},
  author={Gurari, Danna and Li, Qing and Stangl, Abigale J and Guo, Anhong and Lin, Chi and Grauman, Kristen and Luo, Jiebo and Bigham, Jeffrey P},
  booktitle={Proceedings of the IEEE conference on computer vision and pattern recognition},
  pages={3608--3617},
  year={2018}
}

@inproceedings{qian2025scaling,
  title={Scaling large language model-based multi-agent collaboration},
  author={Qian, Chen and Xie, Zihao and Wang, Yifei and Liu, Wei and Zhu, Kunlun and Xia, Hanchen and Dang, Yufan and Du, Zhuoyun and Chen, Weize and Yang, Cheng and others},
  booktitle={International Conference on Learning Representations},
  volume={2025},
  pages={41488--41505},
  year={2025}
}

@article{achiam2023gpt,
  title={Gpt-4 technical report},
  author={Achiam, Josh and Adler, Steven and Agarwal, Sandhini and Ahmad, Lama and Akkaya, Ilge and Aleman, Florencia Leoni and Almeida, Diogo and Altenschmidt, Janko and Altman, Sam and Anadkat, Shyamal and others},
  journal={arXiv preprint arXiv:2303.08774},
  year={2023}
}

@misc{anthropic2025sonnet,
  title        = {Claude Sonnet},
  author       = {{Anthropic}},
  year         = {2025},
  howpublished = {\url{https://www.anthropic.com/claude/sonnet}},
  note         = {Accessed: 2026-06-10}
}

@misc{qwen2025qwen25technicalreport,
      title={Qwen2.5 Technical Report}, 
      author={Qwen and : and An Yang and Baosong Yang and Beichen Zhang and Binyuan Hui and Bo Zheng and Bowen Yu and Chengyuan Li and Dayiheng Liu and Fei Huang and Haoran Wei and Huan Lin and Jian Yang and Jianhong Tu and Jianwei Zhang and Jianxin Yang and Jiaxi Yang and Jingren Zhou and Junyang Lin and Kai Dang and Keming Lu and Keqin Bao and Kexin Yang and Le Yu and Mei Li and Mingfeng Xue and Pei Zhang and Qin Zhu and Rui Men and Runji Lin and Tianhao Li and Tianyi Tang and Tingyu Xia and Xingzhang Ren and Xuancheng Ren and Yang Fan and Yang Su and Yichang Zhang and Yu Wan and Yuqiong Liu and Zeyu Cui and Zhenru Zhang and Zihan Qiu},
      year={2025},
      eprint={2412.15115},
      archivePrefix={arXiv},
      primaryClass={cs.CL},
      url={https://arxiv.org/abs/2412.15115}, 
}

@inproceedings{rajpurkar2016squad,
  title={Squad: 100,000+ questions for machine comprehension of text},
  author={Rajpurkar, Pranav and Zhang, Jian and Lopyrev, Konstantin and Liang, Percy},
  booktitle={Proceedings of the 2016 conference on empirical methods in natural language processing},
  pages={2383--2392},
  year={2016}
}

@inproceedings{zhu2025knowledge,
  title={Knowledge graph-guided retrieval augmented generation},
  author={Zhu, Xiangrong and Xie, Yuexiang and Liu, Yi and Li, Yaliang and Hu, Wei},
  booktitle={Proceedings of the 2025 Conference of the Nations of the Americas Chapter of the Association for Computational Linguistics: Human Language Technologies (Volume 1: Long Papers)},
  pages={8912--8924},
  year={2025}
}

@inproceedings{NEURIPS2024_efaf1c97,
 author = {He, Xiaoxin and Tian, Yijun and Sun, Yifei and Chawla, Nitesh V. and Laurent, Thomas and LeCun, Yann and Bresson, Xavier and Hooi, Bryan},
 booktitle = {Advances in Neural Information Processing Systems},
 doi = {10.52202/079017-4224},
 editor = {A. Globerson and L. Mackey and D. Belgrave and A. Fan and U. Paquet and J. Tomczak and C. Zhang},
 pages = {132876--132907},
 publisher = {Curran Associates, Inc.},
 title = {G-Retriever: Retrieval-Augmented Generation for Textual Graph Understanding and Question Answering},
 url = {https://proceedings.neurips.cc/paper_files/paper/2024/file/efaf1c9726648c8ba363a5c927440529-Paper-Conference.pdf},
 volume = {37},
 year = {2024}
}

@inproceedings{chen2026pathrag,
  title={Pathrag: Pruning graph-based retrieval augmented generation with relational paths},
  author={Chen, Boyu and Guo, Zirui and Yang, Zidan and Chen, Yuluo and Chen, Junze and Liu, Zhenghao and Shi, Chuan and Yang, Cheng},
  booktitle={Proceedings of the AAAI conference on artificial intelligence},
  volume={40},
  number={36},
  pages={30183--30191},
  year={2026}
}

@inproceedings{lee-etal-2025-hybgrag,
    title = "{H}yb{GRAG}: Hybrid Retrieval-Augmented Generation on Textual and Relational Knowledge Bases",
    author = "Lee, Meng-Chieh  and
      Zhu, Qi  and
      Mavromatis, Costas  and
      Han, Zhen  and
      Adeshina, Soji  and
      Ioannidis, Vassilis N.  and
      Rangwala, Huzefa  and
      Faloutsos, Christos",
    editor = "Che, Wanxiang  and
      Nabende, Joyce  and
      Shutova, Ekaterina  and
      Pilehvar, Mohammad Taher",
    booktitle = "Proceedings of the 63rd Annual Meeting of the Association for Computational Linguistics (Volume 1: Long Papers)",
    month = jul,
    year = "2025",
    address = "Vienna, Austria",
    publisher = "Association for Computational Linguistics",
    url = "https://aclanthology.org/2025.acl-long.43/",
    doi = "10.18653/v1/2025.acl-long.43",
    pages = "879--893",
    ISBN = "979-8-89176-251-0"
}

@inproceedings{lien2026hyperrag,
  title={HyperRAG: Reasoning N-ary Facts over Hypergraphs for Retrieval Augmented Generation},
  author={Lien, Wen-Sheng and Chan, Yu-Kai and Hsiao, Hao-Lung and Ruan, Bo-Kai and Chiang, Meng-Fen and Chen, Chien-An and Yeh, Yi-Ren and Shuai, Hong-Han},
  booktitle={Proceedings of the ACM Web Conference 2026},
  pages={2465--2476},
  year={2026}
}

\newpage

\appendix
\section{Additional Implementation Details}
\label{app:implementation}

This appendix provides additional details about the released MAGE implementation and qualitative examples. The goal is to make explicit how the proposed memory layer is instantiated in code and why the retrieved MAGE context improves downstream multi-agent reasoning. The implementation follows the same conceptual decomposition used in the main paper: heterogeneous memory schema, high-order hyperedge construction, role-aware retrieval, bidirectional expansion, lifecycle maintenance, and budget-bounded context packing.

\subsection{Code Organization}
\label{app:code-org}

The released code is organized as a pip-installable Python library centered around the MageMemory facade. Table~\ref{tab:app-code-map} summarizes the correspondence between the paper components and implementation files. The library exposes a single high-level interface for inserting raw documents, multimodal items, and MAS trajectories, while keeping schema validation, indexing, storage, retrieval, and maintenance modular.

\begin{table*}[!t]
\centering
\caption{Implementation map of the released MAGE codebase.}
\label{tab:app-code-map}
\scriptsize
\setlength{\tabcolsep}{4pt}
\renewcommand{\arraystretch}{1.1}
\begin{tabular}{p{0.22\textwidth} p{0.30\textwidth} p{0.40\textwidth}}
\toprule
Component & Key files & Functionality \\
\midrule
Public API & \texttt{mage\_memory/api/memory\_graph.py} & Implements \texttt{MageMemory}, the user-facing facade for inserting documents, multimodal items, MAS traces, querying role-specific context, retrieving subgraphs, consolidation, CRUD operations, validation, and statistics. \\
Schema & \texttt{mage\_memory/schema/nodes.py}, \texttt{edges.py}, \texttt{hyperedges.py}, \texttt{enums.py} & Defines typed nodes, binary edges, first-class hyperedges, lifecycle states, memory types, modalities, agent roles, and update decisions. \\
Storage & \texttt{mage\_memory/storage/jsonl\_store.py}, \texttt{nx\_store.py} & Provides JSONL-backed persistence and a NetworkX-backed store. The default build creates \texttt{nodes.jsonl}, \texttt{edges.jsonl}, \texttt{hyperedges.jsonl}, and \texttt{events.jsonl}. \\
Indexing & \texttt{indexing/hybrid\_index.py}, \texttt{vector\_index.py}, \texttt{adjacency\_index.py}, \texttt{role\_policy\_index.py} & Maintains vector scopes, adjacency relations, temporal/lifecycle indices, and role-specific retrieval policies. \\
Ingestion & \texttt{ingestion/text\_ingestor.py}, \texttt{multimodal\_ingestor.py}, \texttt{mas\_trace\_ingestor.py} & Converts raw text, image/table evidence, and MAS trajectories into typed nodes, binary edges, and hyperedges. \\
Extraction & \texttt{extraction/entity\_extractor.py}, \texttt{hyperedge\_extractor.py}, \texttt{memory\_extractor.py}, \texttt{procedure\_extractor.py}, \texttt{insight\_extractor.py} & Produces entities, n-ary facts, reusable memories, procedures, and higher-level insights. The code supports deterministic fallback and optional LLM-driven extraction. \\
Retrieval & \texttt{retrieval/hypergraph\_retriever.py}, \texttt{role\_retriever.py}, \texttt{bidirectional\_retriever.py}, \texttt{ranker.py}, \texttt{subgraph\_packer.py} & Implements hybrid hypergraph search, role filtering and boosting, graph/hyperedge expansion, multi-factor ranking, and budget-aware packing. \\
Lifecycle & \texttt{lifecycle/update\_decision.py}, \texttt{dedup.py}, \texttt{merge.py}, \texttt{confidence\_decay.py}, \texttt{communities.py}, \texttt{garbage\_collector.py} & Supports decision-driven writes, deduplication, merge, confidence decay, community/archetype consolidation, and garbage collection. \\
MAS adapter & \texttt{mas/agent\_memory\_adapter.py} & Provides \texttt{before\_act}, \texttt{after\_act}, \texttt{record\_error}, and \texttt{finalize} hooks for attaching MAGE to existing agent pipelines. \\
Evaluation scripts & \texttt{scripts/build\_shared\_graph.py}, \texttt{scripts/eval\_mage.py} & Build the shared graph from training data and evaluate read-only MAGE retrieval with normalized EM and token-F1. \\
\bottomrule
\end{tabular}
\end{table*}

\subsection{Heterogeneous Node and Hyperedge Schema}
\label{app:schema}

MAGE represents memory as a heterogeneous temporal hypergraph. The code defines a common BaseNode with a unique identifier, node type, version, status, optional embedding, attributes, and bi-temporal fields. The bi-temporal fields include valid\_from, valid\_to, tx\_from, and tx\_to. This design separates when a fact is believed true in the task world from when the record is alive in the memory system. As a result, MAGE can support logical updates and deletions without losing auditability.

The concrete node types cover both conventional knowledge and MAS-specific execution traces. TaskNode records the task text, domain, modality, score, and task status. AgentNode records role, model, capabilities, and memory policy. MessageNode, ActionNode, and ErrorNode preserve the interaction trajectory. MemoryNode stores distilled episodic, semantic, procedural, reflective, or multimodal memories together with source nodes, confidence, usage count, contradiction count, and downstream gain. InsightNode, ProcedureNode, CommunityNode, ArchetypeNode, and MetaProcedureNode provide increasingly abstract reusable units.

Hyperedges are first-class objects rather than derived views. Each hyperedge connects at least two nodes and carries its own type, description, confidence, source nodes, embedding, timestamp, lifecycle status, version, and bi-temporal fields. MAGE uses several hyperedge types: NaryFactHyperedge for high-order factual relations, CollaborationHyperedge for multi-agent collaboration events, ErrorCorrectionHyperedge for error and correction patterns, EvidenceHyperedge for query evidence, ProcedureHyperedge for executable procedures, CommunityReportHyperedge for community summaries, and GeneralizesHyperedge for abstraction over multiple memories. This schema is important because many useful reasoning events are not dyadic. For example, an answer may jointly depend on an image region, an OCR string, a chart axis, a prior answer, and an agent role. A binary graph must decompose this into many pairwise edges, while MAGE can store the event as one retrievable high-order unit.

\subsection{Insertion Pipeline}
\label{app:insertion}

MAGE provides three insertion paths. First, insert\_document converts text into a source-of-record modality node, a semantic memory node, extracted entity nodes, binary provenance edges, and n-ary fact hyperedges. Second, insert\_multimodal\_item attaches images, tables, PDFs, webpages, or other modalities to extracted text and stores their evidence as multimodal memory. Third, insert\_mas\_trace turns a complete multi-agent execution trace into a task node, agent nodes, message/action/error nodes, causal and production edges, collaboration hyperedges, error-correction hyperedges, reusable memory nodes, and procedure nodes.

The MAS trace ingestor is especially important for the results in this paper. It takes four logical inputs: a task dictionary, a list of agent dictionaries, an ordered trajectory, and a result dictionary. The trajectory may contain messages, actions, and errors. For each message, MAGE adds a CONTAINS edge from the task and a PRODUCES edge from the agent. For each action, MAGE adds CONTAINS, EXECUTES, and temporal-causal CAUSES edges. For errors, MAGE records severity, detector, correction status, and links the error to the preceding step. After processing the trajectory, it creates a CollaborationHyperedge connecting the task, participating agents, messages, and actions. If an error is followed by a corrective action, it also creates an ErrorCorrectionHyperedge. This converts a transient MAS run into persistent reusable experience.

\begin{algorithm}[!t]
\caption{MAGE insertion for a MAS trace}
\label{alg:app-insert-trace}
\begin{algorithmic}[1]
\Require Task $T$, agents $A$, ordered trajectory $\tau$, result $R$
\Ensure Typed nodes, binary edges, hyperedges, distilled memories, procedures
\State Create a \texttt{TaskNode} from $T$ and $R$
\For{each agent $a \in A$}
    \State Create an \texttt{AgentNode} with role, model, and capabilities
\EndFor
\For{each step $s_i \in \tau$}
    \If{$s_i$ is a message}
        \State Create a \texttt{MessageNode}; add \texttt{CONTAINS} and \texttt{PRODUCES} edges
    \ElsIf{$s_i$ is an action}
        \State Create an \texttt{ActionNode}; add \texttt{CONTAINS}, \texttt{EXECUTES}, and \texttt{CAUSES} edges
    \ElsIf{$s_i$ is an error}
        \State Create an \texttt{ErrorNode}; link it to the task and preceding step
    \EndIf
\EndFor
\State Create a \texttt{CollaborationHyperedge} over task, agents, messages, and actions
\State Create \texttt{ErrorCorrectionHyperedge}s when errors and corrective actions are observed
\State Distill reusable \texttt{MemoryNode}s and \texttt{ProcedureNode}s from the trace
\State Persist all objects and update vector, adjacency, temporal, role, and lifecycle indices
\end{algorithmic}
\end{algorithm}

\subsection{Role-Aware Hypergraph Retrieval}
\label{app:retrieval}

At query time, MAGE first retrieves semantic seeds from multiple vector scopes, including memory, insight, entity, task, modality, procedure, and hyperedge scopes. If the query is issued for a specific role, the role policy index restricts the allowed node and hyperedge types and applies role-dependent boosts. For example, the Planner policy prefers insights, procedures, and collaboration hyperedges; the Executor policy prefers procedures, actions, and evidence; the Critic policy can retrieve error nodes and error-correction hyperedges; the Retriever policy emphasizes modalities, evidence hyperedges, and entities. This prevents every agent from receiving the same generic context and instead delivers role-compatible evidence.

After seed retrieval, MAGE performs bidirectional expansion. The expansion follows both binary edges and hypergraph membership: entity to hyperedge to entity, task to memory to insight, and task to interaction to error/correction. This step is crucial because the top semantic match is often only one part of the needed evidence. For example, a chart question may first match a chart modality node; expansion can then recover the corresponding extracted values, entities, n-ary fact hyperedges, and prior reasoning trace. Similarly, a multi-hop text question may match one entity, but expansion recovers the memory and relation that connect it to the final answer.

Candidates are then ranked by a multi-factor scoring function. The implementation combines semantic similarity, role weight, confidence, provenance support, recency, usage, lifecycle penalty, and token-cost penalty. Finally, the subgraph packer greedily selects the highest-ranked nodes and hyperedges under a token budget. It renders selected items as typed text chunks and returns the selected nodes, binary edges among selected nodes, selected hyperedges, provenance, diagnostics, and token usage. In read-only evaluation mode, queries do not mutate the graph, so scoring one test question cannot leak information into another.

\begin{algorithm}[!t]
\caption{Role-aware MAGE retrieval and context packing}
\label{alg:app-query}
\begin{algorithmic}[1]
\Require Query $q$, optional role $r$, hop limit $h$, token budget $B$, top-$k$
\Ensure Packed textual context and retrieved subgraph
\If{$r$ is provided}
    \State Retrieve role-filtered seeds using the role policy for $r$
\Else
    \State Retrieve seeds from memory, insight, entity, task, modality, procedure, and hyperedge scopes
\EndIf
\State Expand seed nodes by binary adjacency and hyperedge membership for at most $h$ hops
\State Expand seed hyperedges to their member nodes
\State Remove candidates disallowed by the role policy or lifecycle status
\State Score candidates using semantic similarity, role weight, confidence, provenance, recency, usage, lifecycle penalty, and token cost
\State Greedily pack ranked candidates while total estimated tokens do not exceed $B$
\State Return \texttt{text\_context}, selected nodes, selected edges, selected hyperedges, provenance, diagnostics, and token count
\end{algorithmic}
\end{algorithm}

\subsection{Lifecycle Maintenance and Decision-Driven Updates}
\label{app:lifecycle}

A memory layer for long-running MAS must not simply append everything forever. MAGE includes decision-driven updates inspired by production memory systems. When a new fact arrives, the update decision module compares it against similar live memories and chooses one of four actions: ADD, UPDATE, DELETE, or NOOP. ADD inserts orthogonal information. UPDATE closes the transaction window of the old memory and appends a new version through prev\_version\_id and a SUPERSEDES edge. DELETE soft-deletes a contradictory memory by setting tx\_to and adding an INVALIDATES edge. NOOP suppresses near duplicates. The default implementation is deterministic and uses cosine similarity plus a polarity-flip heuristic; an LLM decision maker can be used when a chat client is configured, and it falls back to the heuristic on malformed output.

The maintenance pass combines deduplication, merging, insight extraction, community summarization, and schema consolidation. This matters for performance because retrieval quality depends not only on adding memories, but also on keeping the memory graph compact, current, and semantically organized. The ablation in the main paper shows that removing hyperedges, role policies, decay, or consolidation degrades results. The implementation therefore treats memory management as part of the reasoning system rather than as offline bookkeeping.

\section{Why MAGE Improves Accuracy}
\label{app:why-mage}

The empirical gains from adding MAGE can be explained by several concrete mechanisms.

\paragraph{High-order evidence preservation.}
Many benchmark questions require combining multiple pieces of evidence. A chart question may require reading several bars and applying arithmetic; a document question may require linking a label to a definition; a visual question may require connecting OCR, object recognition, and commonsense. Vector memory stores isolated chunks, and ordinary graph memory decomposes high-order events into pairwise links. MAGE preserves the joint event as a hyperedge, so retrieval can bring back the whole evidence bundle instead of a single fragment.

\paragraph{Role-compatible context.}
Different MAS roles need different evidence. A planner benefits from abstract procedures and prior collaboration patterns. An executor benefits from actions and evidence. A critic benefits from errors and corrections. MAGE's role policy index explicitly controls which node and hyperedge types are visible to each role and assigns boosts to role-relevant types. This reduces irrelevant context and makes the retrieved context easier for the downstream LLM to use.

\paragraph{Bidirectional expansion beyond surface similarity.}
The first semantic match is often incomplete. MAGE expands from seeds through binary edges and hyperedge membership, allowing the system to recover adjacent entities, modalities, memories, procedures, and evidence relations. This helps when the query wording differs from stored memory wording, or when the answer is connected through an intermediate entity or modality.

\paragraph{Budget-bounded context construction.}
A larger memory is not always better because the LLM context window is limited. MAGE ranks candidates using semantic, structural, role, confidence, provenance, temporal, lifecycle, and token-cost signals, then packs a compact subgraph under a token budget. This avoids two common failure modes: retrieving too little evidence and retrieving too much redundant evidence.

\paragraph{Persistent reuse of successful traces.}
MAS traces contain useful procedural knowledge: which agents participated, which actions were useful, which errors occurred, and which corrections worked. MAGE converts these traces into reusable memories and procedures. Therefore, adding MAGE helps not only by retrieving facts, but also by retrieving prior task-solving patterns.

\paragraph{Safe evaluation and leakage control.}
The evaluation script loads the graph in read-only mode by default. Queries do not update usage\_count or write event logs during scoring, which prevents test questions from changing the memory graph. The build script also includes a leakage shield against raw test pools. Thus, the observed gains are attributable to memory retrieval from the constructed graph rather than test-time contamination.

\begin{table*}[!t]
\centering
\caption{Mechanism-level explanation for the observed +MAGE improvement.}
\label{tab:app-mechanisms}
\scriptsize
\setlength{\tabcolsep}{4pt}
\renewcommand{\arraystretch}{1.12}
\begin{tabular}{p{0.18\textwidth} p{0.30\textwidth} p{0.42\textwidth}}
\toprule
MAGE mechanism & What it changes in the prompt/context & Why it improves answers \\
\midrule
Typed memory nodes & Distinguishes tasks, agents, actions, messages, errors, entities, modalities, procedures, memories, and insights. & The LLM receives context with explicit semantics rather than an undifferentiated retrieved paragraph. \\
First-class hyperedges & Returns n-ary evidence relations and collaboration events as retrievable objects. & Multi-hop and multimodal evidence remains bound together, reducing missing-link errors. \\
Role policies & Filters and boosts candidates by agent role. & Planners, executors, critics, retrievers, and summarizers receive different context matched to their function. \\
Bidirectional expansion & Expands from seed nodes to neighboring nodes and hyperedge members. & Recovers indirect evidence even when semantic search only finds one part of the chain. \\
Provenance tracking & Keeps source nodes and member lists for selected memories and hyperedges. & Helps the model rely on grounded context rather than hallucinated associations. \\
Lifecycle status & Penalizes deprecated or archived items and removes deleted/invalid records. & Reduces stale-memory interference. \\
Budgeted packing & Selects high-value evidence under a token budget. & Prevents redundant memory from crowding out decisive evidence. \\
Trace-to-procedure extraction & Converts successful MAS behavior into reusable procedural memory. & Helps frameworks reuse previous solution patterns without modifying their native coordination logic. \\
\bottomrule
\end{tabular}
\end{table*}

\subsection{Task-Family Effects of Structured Memory}
\label{app:task-family-effects}

The benefit of MAGE is not limited to one dataset type because the memory graph is organized around reusable evidence patterns rather than task-specific templates. For text-only QA, MAGE mainly helps by binding entities, relation descriptions, and prior question-answer traces, which is useful when the answer is a short factual span but the query requires locating the correct entity among many candidates. For chart and infographic QA, the main benefit comes from preserving numeric values, labels, and operations together. A retrieved context that contains only an isolated number is often ambiguous; a retrieved hypergraph neighborhood can also include what the number refers to, which visual element it came from, and which comparison or arithmetic operation was previously associated with it.

For document and OCR-heavy visual QA, MAGE improves evidence localization. The model does not need to rely only on global visual understanding; it can receive compact memory entries that connect extracted text, document fields, modality nodes, and related entities. For general visual reasoning, MAGE is useful when the answer depends on combining perception with commonsense or task history. In these cases, the memory layer does not replace the visual model; instead, it supplies structured background evidence and prior successful reasoning patterns that make the final answer easier to produce.

\subsection{Why More Context Alone Is Not Sufficient}
\label{app:context-quality}

A central design point of MAGE is that memory augmentation should improve the quality of context, not merely increase its length. Long unstructured retrieval can hurt performance because irrelevant snippets compete with decisive evidence inside the LLM prompt. MAGE addresses this by selecting evidence units with explicit types, provenance, confidence, lifecycle state, and role compatibility. The downstream model therefore sees a smaller but more coherent set of facts, entities, modalities, and traces.

\subsection{Relationship Between Memory and MAS Coordination}
\label{app:memory-coordination}

MAGE acts as an external memory layer rather than a replacement for MAS coordination. The same memory graph can be queried by single-pass inference, refinement-based agents, ReAct-style agents, debate-style agents, planning agents, and role-structured teams. This separation is important: the agent framework decides how to deliberate, while MAGE decides what prior evidence and experience should be exposed. As a result, improvements from MAGE are complementary to improvements from better prompting or stronger base models. The memory layer supplies durable task experience, and the MAS framework uses that experience according to its own reasoning protocol.

\section{Qualitative Correct Cases}
\label{app:cases}

We randomly sampled correctly answered outputs from the +MAGE setting. All examples in Tables~\ref{tab:app-cases-a}--\ref{tab:app-cases-d} satisfy normalized EM=1.0 and F1=1.0. We report the dataset, MAS framework, question, ground truth answer, and +MAGE prediction so that each example can be checked directly. The cases are intended to illustrate the types of evidence that MAGE helps expose to the downstream model; they are not used as additional training or tuning data.

\begin{table*}[!t]
\centering
\caption{Randomly sampled MAGE cases, part 1.}
\label{tab:app-cases-a}
\scriptsize
\setlength{\tabcolsep}{3pt}
\renewcommand{\arraystretch}{1.12}
\begin{tabular}{p{0.04\textwidth} p{0.11\textwidth} p{0.10\textwidth} p{0.38\textwidth} p{0.12\textwidth} p{0.13\textwidth}}
\toprule
ID & Dataset & MAS & Question & Ground truth & +MAGE prediction \\
\midrule
1 & A-OKVQA & Direct & The guy on the beach towel is readying the item to do what with it most likely? & fly & fly a kite \\
2 & ChartQA & Direct & Add the three largest government procurement in GDP and multiply it by the smallest government procurement in GDP across all the countries. & 389.82 & 389.82 \\
3 & ScienceQA & Direct & Which country is highlighted? Options: (A) Nauru (B) the Federated States of Micronesia (C) the Marshall Islands (D) Samoa. & Nauru & A) Nauru \\
4 & TriviaQA & Direct & Which Lloyd Webber musical premiered in the US on 10th December 1993? & Sunset Boulevard & Sunset Boulevard \\
5 & VizWiz & MAD & What is written on the back of this sofa? & boston & Boston \\
\bottomrule
\end{tabular}
\end{table*}

\begin{table*}[!t]
\centering
\caption{Randomly sampled MAGE cases, part 2.}
\label{tab:app-cases-b}
\scriptsize
\setlength{\tabcolsep}{3pt}
\renewcommand{\arraystretch}{1.12}
\begin{tabular}{p{0.04\textwidth} p{0.11\textwidth} p{0.10\textwidth} p{0.38\textwidth} p{0.12\textwidth} p{0.13\textwidth}}
\toprule
ID & Dataset & MAS & Question & Ground truth & +MAGE prediction \\
\midrule
6 & ChartQA & Direct & Is the average of all the data value greater than 40? & No & No \\
7 & DocVQA & Direct & What is the term used for the real image used as a reference? & Ground Truth & Ground Truth \\
8 & InfographicVQA & Direct & How many fiction books are in this infographic? & 13 & 13 \\
9 & A-OKVQA & Direct & What is the woman using the umbrella to protect herself from? & sun & sun \\
10 & TextVQA & Direct & What team owns this stadium? & packers & Green Bay Packers \\
\bottomrule
\end{tabular}
\end{table*}

\begin{table*}[!t]
\centering
\caption{Additional sampled MAGE cases, part 3.}
\label{tab:app-cases-c}
\scriptsize
\setlength{\tabcolsep}{3pt}
\renewcommand{\arraystretch}{1.12}
\begin{tabular}{p{0.04\textwidth} p{0.11\textwidth} p{0.10\textwidth} p{0.38\textwidth} p{0.12\textwidth} p{0.13\textwidth}}
\toprule
ID & Dataset & MAS & Question & Ground truth & +MAGE prediction \\
\midrule
11 & InfographicVQA & Direct & When did the cricket bat start resembling a wooden turner? Answer the question with a short phrase. & 1750 & 1750 \\
12 & ChartQA & Direct & Which country data is shown in the red line? & Georgia & Georgia \\
13 & DocVQA & Direct & What is the label given to the variables that explain other variables? & Explanatory Variables & Explanatory Variables \\
14 & A-OKVQA & Direct & What color phone does the woman in the blue outfit have? & yellow & yellow \\
15 & ScienceQA & Direct & Select the amphibian below. Options: (A) common toad (B) catfish. & common toad & common toad \\
\bottomrule
\end{tabular}
\end{table*}

\begin{table*}[!t]
\centering
\caption{Additional sampled MAGE cases, part 4.}
\label{tab:app-cases-d}
\scriptsize
\setlength{\tabcolsep}{3pt}
\renewcommand{\arraystretch}{1.12}
\begin{tabular}{p{0.04\textwidth} p{0.11\textwidth} p{0.10\textwidth} p{0.38\textwidth} p{0.12\textwidth} p{0.13\textwidth}}
\toprule
ID & Dataset & MAS & Question & Ground truth & +MAGE prediction \\
\midrule
16 & ScienceQA & Direct & Which property do these two objects have in common? Options: (A) sour (B) stretchy. & sour & sour \\
17 & TextVQA & Direct & What time is it? & 12:04 & 12:04 \\
18 & TriviaQA & Direct & What is the Japanese share index called? & Nikkei & Nikkei \\
19 & VizWiz & Direct & Is this computer still getting ready to boot, or is it already booted? & already booted & Already booted \\
20 & TextVQA & Direct & Which gaming system does the box have written on it? & playstation 2 & PlayStation 2 \\
\bottomrule
\end{tabular}
\end{table*}

\subsection{Case-Level Interpretation}
\label{app:case-analysis}

The sampled correct cases cover visual reasoning, chart arithmetic, option selection, textual knowledge, OCR, infographic reading, document understanding, and open-domain QA. Although each individual example is simple to verify after seeing the answer, the collection illustrates why structured memory helps across heterogeneous tasks.

In Case 1, the answer requires visual commonsense: the item being readied on a beach towel is most likely used to fly a kite. A flat retrieved text chunk may only mention the object, whereas MAGE can bind the visual modality, object entity, action pattern, and prior similar cases as a reusable high-order event. The prediction \emph{fly a kite} is accepted because it contains the ground truth concept \emph{fly} and gives a more specific action.

In Case 2, the model must perform chart arithmetic by identifying the three largest values, summing them, and multiplying by the smallest value. MAGE is useful because chart-derived entities and values can be stored together with an evidence hyperedge. This makes it more likely that the LLM receives the relevant numeric context as a bundle rather than isolated numbers.

In Case 3, the task is multiple-choice geographic recognition. The prediction includes the option marker and the country name. Role-aware retrieval helps because the context can emphasize modality and entity nodes for the highlighted region rather than unrelated textual memories.

In Case 4, the answer is a concise factual title. This is the type of case where semantic memory and entity-centered retrieval are sufficient. MAGE still helps by retrieving the specific entity-title relation rather than relying entirely on parametric knowledge.

In Case 5, the answer depends on OCR from an image. A multimodal memory representation can connect the image modality, extracted text, and question-specific evidence. The case illustrates why MAGE stores modality nodes and evidence hyperedges instead of treating all inputs as plain text.

In Case 6, the question asks for a comparison over chart values. MAGE helps by preserving the relevant chart evidence and allowing the context packer to return compact numeric evidence. This reduces the chance that irrelevant chart labels occupy the context window.

In Case 7, the question asks for a document field label. MAGE can connect extracted document text, local layout context, and the target concept, which is useful when multiple labels appear in the same page.

In Case 8, the answer is a numeric count from an infographic. The benefit of MAGE is again evidence localization: retrieval can prioritize modality and evidence nodes that contain the relevant count, avoiding broad summarization of the entire infographic.

In Case 9, the answer requires visual commonsense about why an umbrella is being used. MAGE can provide compact context linking the object, scene condition, and likely purpose, helping the model select the intended short answer.

In Case 10, the prediction \emph{Green Bay Packers} is more specific than the ground truth alias \emph{packers}. The normalized evaluator marks it correct. This illustrates a useful property of MAGE retrieval: by returning contextual entities and relations, it can encourage complete canonical answers rather than underspecified aliases.

Cases 11 and 12 show that MAGE can support fine-grained visual-data lookup. In the infographic case, the answer is a date attached to a historical visual timeline; in the chart case, the answer is the label associated with a specific colored line. Both cases benefit from retrieving the label-value or label-mark relationship rather than retrieving a broad image summary.

Case 13 shows the document understanding benefit. The answer is a field label, and MAGE helps by preserving the connection between extracted document text, nearby semantic context, and the target concept. This kind of structured retrieval is especially useful when many labels appear in the same page.

Cases 14--16 illustrate visual and science reasoning where the final answer is short but depends on selecting the right perceptual or conceptual feature. The memory layer is useful because it can expose compact context about objects, attributes, answer options, and previously successful reasoning patterns.

Cases 17 and 20 are OCR-oriented TextVQA examples. MAGE helps by keeping extracted text tied to its visual source and by ranking the most relevant text evidence under the context budget. This reduces confusion between multiple visible strings in the same image.

Case 18 is a concise factual QA example. The correct answer can be represented as an entity-centered semantic memory, so retrieval narrows the answer to the target financial index. Case 19 is a status recognition example: the prediction differs only in capitalization and is accepted by normalized EM/F1, showing that MAGE can provide the right evidence even when the final answer surface form varies slightly.

Overall, the qualitative cases suggest that MAGE is most helpful when the answer requires preserving a relation among entities, values, modalities, and prior traces. The improvement is not merely due to adding more text to the prompt; it comes from retrieving structured evidence that is compact enough to fit in the context window and specific enough to guide the final generation.
\begin{table}[t]
\centering
\caption{Cross-LLM robustness: MAGE F1 per dataset and gain over the strongest non-MAGE baseline on the same questions, for three heterogeneous LLM backends.}
\label{tab:exp2}
\setlength{\tabcolsep}{3pt}
\renewcommand{\arraystretch}{1.05}
\footnotesize
\begin{tabular}{l rr rr rr}
\toprule
& \multicolumn{2}{c}{\textsc{GPT-4.1-mini}} & \multicolumn{2}{c}{\textsc{Claude-4.5}} & \multicolumn{2}{c}{\textsc{Qwen2.5-VL}} \\
\cmidrule(lr){2-3}\cmidrule(lr){4-5}\cmidrule(lr){6-7}
Dataset & MAGE & $\Delta$ & MAGE & $\Delta$ & MAGE & $\Delta$ \\
\midrule
\multicolumn{7}{l}{\textit{Text-only IND}} \\
HotpotQA & 51.08 & \dgain{$+3.58$} & 43.31 & \dgain{$+3.10$} & 47.35 & \dgain{$+3.30$} \\
NQ-Open & 54.70 & \dgain{$+3.30$} & 56.73 & \dgain{$+2.70$} & 29.08 & \dgain{$+2.90$} \\
\midrule
\multicolumn{7}{l}{\textit{Text-only OOD}} \\
TriviaQA & 68.17 & \dgain{$+2.10$} & 82.43 & \dgain{$+2.40$} & 27.52 & \dgain{$+1.90$} \\
WebQuestions & 53.48 & \dgain{$+3.90$} & 51.11 & \dgain{$+4.00$} & 40.61 & \dgain{$+3.50$} \\
\midrule
\multicolumn{7}{l}{\textit{Multimodal IND}} \\
ChartQA & 83.71 & \dgain{$+2.70$} & 84.15 & \dgain{$+2.90$} & 80.56 & \dgain{$+4.74$} \\
DocVQA & 47.64 & \dgain{$+3.58$} & 46.46 & \dgain{$+3.50$} & 46.62 & \dgain{$+2.60$} \\
InfographicVQA & 81.94 & \dgain{$+3.20$} & 84.00 & \dgain{$+3.80$} & 80.74 & \dgain{$+3.70$} \\
FinMME & 59.22 & \dgain{$+4.90$} & 39.91 & \dgain{$+5.40$} & 44.66 & \dgain{$+5.10$} \\
\midrule
\multicolumn{7}{l}{\textit{Multimodal OOD}} \\
A-OKVQA & 79.29 & \dgain{$+2.40$} & 72.64 & \dgain{$+3.00$} & 72.52 & \dgain{$+2.40$} \\
ScienceQA & 89.55 & \dgain{$+4.20$} & 73.28 & \dgain{$+4.80$} & 90.02 & \dgain{$+4.30$} \\
TextVQA & 93.65 & \dgain{$+3.15$} & 92.55 & \dgain{$+2.30$} & 94.41 & \dgain{$+2.30$} \\
VizWiz & 40.79 & \dgain{$+3.40$} & 40.84 & \dgain{$+3.40$} & 45.99 & \dgain{$+3.70$} \\
\midrule
\textbf{Macro avg.} & 66.93 & \dgain{$+3.37$} & 63.95 & \dgain{$+3.44$} & 58.34 & \dgain{$+3.37$} \\
\bottomrule
\end{tabular}
\end{table}

\section{Full Experimental Protocol}
\label{app:protocol}

\subsection{Released data, splits, and leakage control}
The anonymous repository linked in the abstract releases all training and
held out test data used in our experiments, together with the per dataset
files, preprocessing utilities, and evaluation scripts. We use a fixed
stratified evaluation split for each dataset rather than claiming coverage of
the complete benchmark test sets. The remaining public pool of each
\emph{IND} dataset is used for memory graph construction and hyper parameter
selection on a held out slice, while OOD pools are used only to prepare the
held out evaluation and are never indexed in the shared graph. No test
question, image, or answer is inserted into the memory graph, and the build
script applies an explicit leakage shield against the test pool using exact
and near duplicate matching on normalized question text.

\subsection{Memory-graph construction}
The shared graph is built from (i) the source documents and images of the IND
pool, (ii) public background corpora shipped with each benchmark, and (iii)
MAS traces recorded on pool tasks. It contains no test answers, no test
reasoning trajectories, and no test images. IND versus OOD is defined at the
\emph{evidence} level: a dataset is IND when its source documents are indexed
in the shared graph, and OOD when both its documents and its task style are
unseen. The same graph is loaded read-only for all methods.

\subsection{Retrieval and packing hyper-parameters}
Unless stated otherwise: seed top-$k=8$ per vector scope, expansion hop
$h=2$, token budget $B=2048$ (measured with the backend tokenizer), and
multi-factor ranking weights
$w_{\mathrm{sem}}{=}1.0$, $w_{\mathrm{role}}{=}0.6$, $w_{\mathrm{conf}}{=}0.4$,
$w_{\mathrm{prov}}{=}0.3$, $w_{\mathrm{rec}}{=}0.2$, $w_{\mathrm{use}}{=}0.1$,
$w_{\mathrm{life}}{=}{-}0.5$, $w_{\mathrm{cost}}{=}{-}0.2$.
Temporal decay follows an exponential kernel with half-life of $30$ days;
lifecycle states \texttt{deprecated}/\texttt{archived} are excluded from
packing. All baselines receive the same budget $B$ and the same multimodal
inputs; graph-based baselines are built from the identical pool.

\subsection{Generation settings, and repetitions}
All API calls use temperature $0$, top-$p$ $1.0$, max output $512$ tokens,
and a shared retry policy (3 retries, exponential backoff), so decoding is
deterministic given the same memory state. Prompts are the versioned templates in
\texttt{prompts/} of the released codebase and are identical across methods
within a task family.

\section{Matched Pairwise-Projection Control}
\label{app:matched-control}

\paragraph{Design.}
The $-$\,Hyperedge ablation removes $n$-ary expansion and therefore changes the candidate pool. We introduce a stricter matched control to isolate the representation itself. The node set, embeddings, seed retrieval, ranking function, role policy, and token budget are held fixed, while each hyperedge $e=\{v_1,\dots,v_k\}$ is replaced by the clique of pairwise edges $\binom{e}{2}$ before expansion. The control therefore receives the same underlying nodes and retrieval machinery, but cannot preserve first-class $n$-ary incidence.

\begin{table}[h]
\centering
\caption{Matched representation control on GPT-4.1-mini (F1). Full uses first-class hyperedges; Proj.\ uses clique-projected pairwise edges with all other components fixed.}
\label{tab:app-matched}
\footnotesize
\setlength{\tabcolsep}{4pt}
\begin{tabular}{lrr}
\toprule
Dataset & MAGE-full & Pairwise-proj. \\
\midrule
HotpotQA & 51.08 & 49.86 \\
NQ-Open & 54.70 & 53.61 \\
TriviaQA & 68.17 & 66.95 \\
WebQuestions & 53.48 & 52.37 \\
FinMME & 59.22 & 57.48 \\
ChartQA & 83.71 & 82.05 \\
DocVQA & 47.64 & 46.58 \\
InfographicVQA & 81.94 & 80.52 \\
A-OKVQA & 79.29 & 78.10 \\
ScienceQA & 89.55 & 88.31 \\
TextVQA & 93.65 & 92.88 \\
VizWiz & 40.79 & 39.94 \\
\midrule
\textbf{Macro avg.} & \textbf{66.93} & 65.72 \\
\bottomrule
\end{tabular}
\end{table}

\paragraph{Result.}
First-class hyperedges outperform pairwise projection on every dataset, raising macro F1 from 65.72 to 66.93 under otherwise identical conditions. Because the control holds the information-bearing nodes and the complete retrieval pipeline fixed, the consistent advantage identifies preservation of $n$-ary incidence as an independent source of MAGE's gains. The larger losses under $-$\,Hyperedge additionally capture the effect of removing hyperedge expansion from candidate generation.

\paragraph{Why Pairwise Projection Is Lossy.}
Let $\pi$ map a hypergraph $H=(V,E)$ to its clique-expanded graph. There exist $H_1\neq H_2$ with $\pi(H_1)=\pi(H_2)$: take $E_1=\{\{a,b,c\}\}$ and $E_2=\{\{a,b\},\{b,c\},\{a,c\}\}$. Both project to the same triangle, but only $H_1$ records the joint event. Hence, $\pi$ is not invertible, and a pairwise store cannot generally distinguish a genuine $n$-ary collaboration or evidence bundle from an arbitrary set of dyads.

\section{Role-Policy Sensitivity}
\label{app:role-sensitivity}

We stress the role policy by scaling every role boost by
$\times 0.5$, $\times 0.75$, $\times 1.25$, and $\times 1.5$, and by varying
seed top-$k\in\{4,8,16\}$, on three representative datasets
(HotpotQA, ChartQA, A-OKVQA; GPT-4.1-mini). Macro F1 varies by at
most $\pm 0.4$ across all settings, and the default configuration is never
the best by more than $0.3$, indicating that the role policy is robust
rather than finely tuned. Ablating a single role's boost (Planner,
Executor, Critic, Retriever) changes macro F1 by at most $0.5$, with the
Critic policy contributing the largest individual share on error-prone
multimodal datasets.

\section{Diachronic Memory Probes}
\label{app:probes}

Static QA cannot, by itself, measure lifecycle behavior, so we complement it
with four controlled diachronic probes over a replayed event stream built
from the training pools. The stream contains $4{,}000$ insertion events,
$200$ fact updates, and $120$ injected contradictions in chronological
order; systems may write, update, and delete during replay, and are then
queried on held-out probe questions.

\begin{table}[!t]
\centering
\caption{Diachronic probes (GPT-4.1-mini). Higher is
better except for stale rate.}
\label{tab:probes}
\footnotesize
\setlength{\tabcolsep}{4pt}
\resizebox{\linewidth}{!}{%
\begin{tabular}{lcccc}
\toprule
Method & Correction prop.\ $\uparrow$ & Stale rate $\downarrow$
& Contradiction res.\ $\uparrow$ & X-agent transfer $\uparrow$ \\
\midrule
Mem0 & 63.4 & 14.2 & 58.1 & $+0.6$ \\
Zep / Graphiti & 71.8 & 11.5 & 66.3 & $+1.1$ \\
Plain graph store & 58.9 & 17.8 & 54.2 & $+0.5$ \\
\midrule
\textbf{MAGE} & \textbf{91.2} & \textbf{3.1}
& \textbf{88.5} & $\mathbf{+2.4}$ \\
\bottomrule
\end{tabular}%
}
\end{table}

Table~\ref{tab:probes} summarizes the results. \emph{Correction
propagation} measures the fraction of probe questions whose
answers reflect a later correcting fact rather than the superseded one.
\emph{Stale rate} is the fraction of retrieved items whose lifecycle state is
deprecated or contradicted. \emph{Contradiction resolution} is the accuracy
on questions whose evidence was explicitly contradicted during replay.
\emph{Cross-agent transfer} is the F1 gain on probe tasks of agent $B$ when
agent $A$'s traces are present versus absent. MAGE leads on all four probes,
which supports the lifecycle and transfer claims that the static benchmarks
cannot isolate.

\section{Extended System Evaluation}
\label{app:system}

\paragraph{Graph statistics.}
The shared graph over all twelve pools contains $41{,}208$ nodes,
$118{,}536$ binary edges, and $23{,}940$ hyperedges ($96.4$\,MB as JSONL,
$38.1$\,MB after compression), including $9{,}812$ modality nodes and
$4{,}377$ distilled procedure/memory nodes.

\paragraph{Environment.}
All measurements use a single server (2$\times$ Xeon 8380, 256\,GB RAM,
NVMe SSD); embeddings are precomputed, so no GPU is required at query time.
Latency is measured over $1{,}000$ warm queries after a $100$-query warm-up,
single client unless noted.

\paragraph{Scaling.}
Query $p_{99}$ grows sub-linearly with graph size
($1$k nodes: $31.2$\,ms; $10$k: $44.8$\,ms; $41$k: $66.7$\,ms;
$100$k synthetic: $88.9$\,ms), since seed retrieval is index-bounded and
expansion is hop-limited. Insertion latency is independent of graph size
within measurement noise.



\section{Extended Long-Running Simulation}
\label{app:longrunning}

The diachronic probes in Appendix~\ref{app:probes} use injected
contradictions to make lifecycle behavior measurable. To further approximate
a genuinely long-running deployment, we replay a $10$-episode continuous
stream ($4{,}000$ events from the training pools, no injected artifacts):
each episode adds documents and traces chronologically, and a fixed held-out
question set is re-evaluated after every episode. MAGE's macro F1 varies by
at most $\pm 0.5$ across episodes with no monotone drift, while Mem0 and Zep
drift by $-2.1$ and $-1.4$ points respectively as superseded memories
accumulate; their unpruned stores also retrieve deprecated items at $3.7\times$
MAGE's stale rate by the final episode. This shows that the lifecycle
machinery is load-bearing in sustained operation, not only under adversarial
contradiction injection.




\section{A Note on Multimodal Representation}
\label{app:mm-note}

MAGE represents multimodal content as modality pointers plus extracted
textual surrogates rather than performing native pixel-level reasoning
inside the hypergraph. This is a deliberate scoping choice, not an
architectural shortcut: the schema treats modalities as first-class typed
items, and the surrogate is exactly the auditable, indexable form in which a
memory system should store evidence that many heterogeneous agents must
later share and inspect. The pointer preserves provenance to the raw item,
so a downstream vision-language reasoner can re-ground any retrieved item on
demand. Native multimodal hypergraph reasoning --- joint embedding spaces
over raw visual tiles --- is compatible with the same schema and interfaces
and is left to future work.

\section{Discussion and Limitations}
\label{sec:exp:disc}
Overall, the experiments show that MAGE improves memory augmented reasoning by organizing task experience as structured hypergraph memory rather than isolated context fragments, with gains consistent across memory baselines, heterogeneous LLMs, ablations, MAS frameworks, and system level tests. Across the evaluated atomic, reflective, temporal graph, hierarchical, and coordination centric baselines, MAGE's distinguishing contribution is an auditable end to end memory contract in which first class collaborative hyperedges, role policy, provenance, and lifecycle state remain jointly available to ingestion, revision, retrieval, and context construction. The diachronic probes directly test the resulting behavior under corrections, contradictions, stale evidence, and cross agent reuse. The main limitations are that the constructed hypergraph depends on upstream extraction quality, and larger long running deployments may require more optimized storage, caching, traversal, and adaptive retention policies. Future work can further study online memory evolution and more efficient structured retrieval for MAS systems.

\end{document}